\documentclass[10pt,pre,twocolumn,tightenlines,longbibliography,notitlepage]{revtex4-2}
\pdfoutput=1

\usepackage[textsize=tiny]{todonotes}
\usepackage{xcolor}
\usepackage{verbatim}
\usepackage{graphicx}
\usepackage{natbib}
\usepackage{siunitx}

\newcommand{\kbt}{$k_{\mathrm{B}}T$}
\newcommand*{\Nc}{N_\mathrm{C}}

\begin{document}

\title{Tiling a tubule: How increasing complexity improves the yield of self-limited assembly}

\author{Thomas E. Videb\ae k}
\email{videbaek@brandeis.edu}

\author{Huang Fang, Daichi Hayakawa, Botond Tyukodi, Michael F. Hagan} 
\author{W. Benjamin Rogers}
\email{wrogers@brandeis.edu}
\affiliation{Martin A. Fisher School of Physics, Brandeis University, Waltham, MA, 02453, USA}

\begin{abstract}

The ability to design and synthesize ever more complicated colloidal particles opens the possibility of self-assembling a zoo of complex structures, including those with one or more self-limited length scales. An undesirable feature of systems with self-limited length scales is that thermal fluctuations can lead to the assembly of nearby, off-target states. We investigate strategies for limiting off-target assembly by using multiple types of subunits. Using simulations and energetics calculations, we explore this concept by considering the assembly of tubules built from triangular subunits that bind edge to edge. While in principle, a single type of triangle can assemble into tubules with a monodisperse width distribution, in practice, the finite bending rigidity of the binding sites leads to the formation of off-target structures. To increase the assembly specificity, we introduce tiling rules for assembling tubules from multiple species of triangles. We show that the selectivity of the target structure can be dramatically improved by using multiple species of subunits, and provide a prescription for choosing the minimum number of subunit species required for near-perfect yield. Our approach of increasing the system's complexity to reduce the accessibility of neighboring structures should be generalizable to other systems beyond the self-assembly of tubules.

\end{abstract}

\maketitle

\section{Introduction}

In recent years, new techniques have been developed that allow for greater control over the types of interactions that can be prescribed between nanometer- and micrometer-scale colloidal particles. In particular, valence-limited interactions provided by patchy colloidal particles~\cite{kraft2009colloidal, sacanna2013shaping,yi2013recent,wang2014three,oh2020photo,li2020colloidal,he2021colloidal} or DNA origami~\cite{gerling2015dynamic,wagenbauer2017gigadalton, sigl2021programmable} expand the diversity of structures that can be self-assembled, including complex crystal types, like the kagome~\cite{chen2011directed} or diamond lattices~\cite{he2020colloidal}, as well as non-crystalline structures, like colloidal molecules, one-dimensional filaments, and two-dimensional sheets~\cite{zhang2004self, glotzer2007anisotropy, sciortino2007self, zerrouki2008chiral, sacanna2010lock, chen2011supracolloidal, yan2013colloidal, grunwald2014patterns, bharti2015assembly, mahynski2015grafted,  whitelam2015emergent, wolters2015self, whitelam2016minimal, preisler2017irregular, morphew2018programming, tikhomirov2018triangular, oh2019colloidal, ben2021design, mahynski2021symmetry}. The importance of valence-limited interactions in assembling complex structures is that they constrain the number of contacts per particle and can be used to enforce binding at specific angles. This angular specificity has the additional ability to impart an effective curvature to subunits that can cause the assembly to self-close.

A particularly interesting subset of assemblies enabled by specific, directional interactions is self-limited structures, in which one or more dimensions of the final assembly have a finite extent~\cite{hagan2021equilibrium}. These types of self-limiting architectures are common-place in living systems, with examples ranging from microtubules, which are made up of a pair of subunits~\cite{vanburen2005mechanochemical}, to viral capsids~\cite{caspar1962physical}, which can be constructed from a handful of subunits, to ribosomes, which are fully-addressable structures~\cite{klinge2019ribosome}. While groups have made progress in assembling a variety of self-limiting structures from synthetic colloids~\cite{zerrouki2008chiral, sacanna2010lock, chen2011supracolloidal, yan2013colloidal, bharti2015assembly, wolters2015self, tikhomirov2018triangular, oh2019colloidal}, in almost all cases, the self-limiting dimension is comparable to the size of the individual subunits. Creating structures with a self-limited length scale that is larger than the subunits remains a challenge. One recent success in this regard has been seen in the assembly of icosahedral shells using DNA origami, whereby triangular subunits with specific, valence-limited interactions assembled into a range of shells with varying size~\cite{sigl2021programmable}.

A fundamental consequence of introducing a self-limited length scale that is larger than the size of the constituent parts is that the self-limited length can vary due to thermal fluctuations. These fluctuations can cause the system to access unintended final states, leading to a distribution of assemblies rather than just the single target structure~\cite{cheng2014self}. This behavior has been seen in synthetic systems, such as the assembly of rings from wedge-shaped particles made by DNA origami, which form a distribution of ring sizes~\cite{wagenbauer2017gigadalton}. There, the off-target states occur due to neighboring minima in the free-energy landscape. In general, the off-target states can either be accessed in equilibrium, as in the case of the rings mentioned above~\cite{wagenbauer2017gigadalton} as well as in self-limited, multi-component assemblies~\cite{zeravcic2014size, murugan2015multifarious}, or from kinetic traps, where assembly gets caught in a local minimum at early stages of self-assembly and is unable to relax to a lower free energy.  

In this report, we examine how using multiple species of particles can limit the formation of off-target states by engineering the free-energy landscape of assembly. Specifically, we study the assembly of triangular subunits into cylindrical tubules. Using both simulation and energetic calculations, we explore how the number of off-target tubules grows with the designed target width and the bending rigidity of the subunits. By finding allowed tilings of the plane with multiple species of triangles, we  construct interactions that remove nearby off-target states from the energy landscape. We show that when the periodic length scale associated with a multiple-species tiling becomes comparable to the fluctuations of the self-limited size of the assembly, the target structure assembles with near-perfect yield. This criterion defines a minimum number of subunit species that are needed to guarantee assembly of a prescribed architecture, exemplifying the trade-off between the complexity of the assembly and the distribution of assembly outcomes. Our results provide a route to create tubules with precisely controlled widths and could be extended to other self-limiting architectures.

\begin{figure}[t]
 \centering
 \includegraphics[width=86mm]{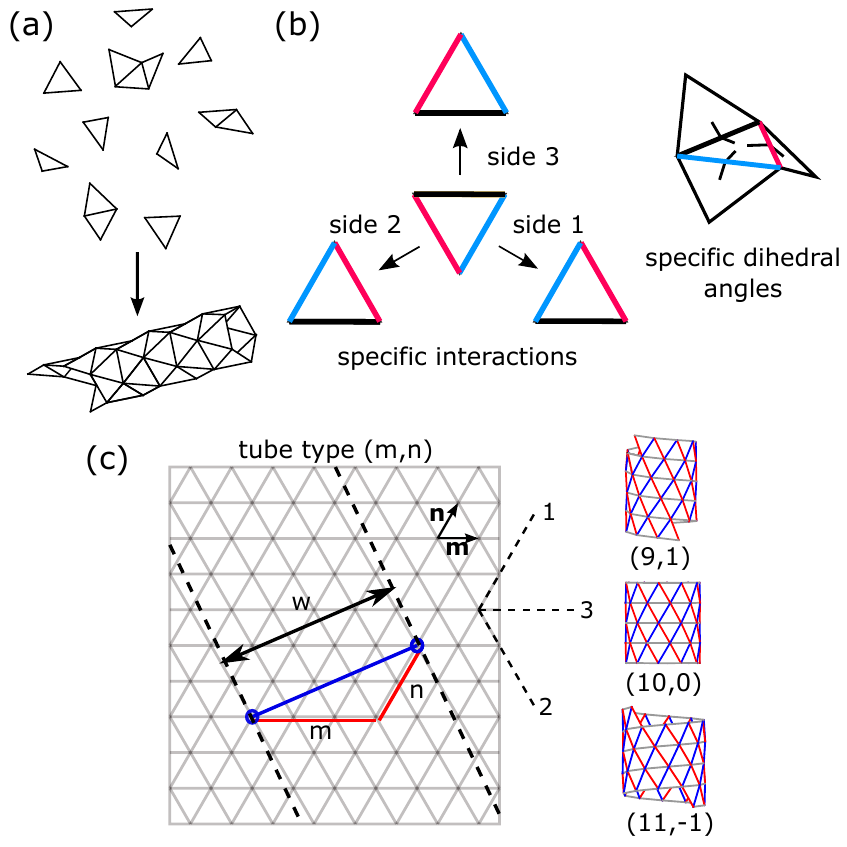}
 \caption{\textit{Design principles of tubule assembly}. (a) Schematic of the self-assembly of triangular subunits into a tubule. (b) Diagram of the specific interactions. Each edge has a specific self-interaction (side 1 binds to side 1, etc.) and has a specified dihedral angle between the edges of two bound subunits. These specific angles act as a form of effective curvature for the subunits. (c) shows how a discrete tubule can be indexed by considering periodic boundary conditions between two vertices of a triangular lattice. The line between the two circled vertices denotes the circumference of the tubule that will form, while the lines labeled by \textit{m} and \textit{n} show the number of steps needed to reach that vertex along lattice directions \textbf{m} and \textbf{n} and has a width \textit{w}. The labeling of (\textit{m,n}) will be used to denote the type of tubule that forms, some examples of which are shown for (9,1), (10,0), and (11,-1). }
 \label{fig:fig1}
\end{figure}

\section{Self-assembly of tubules from triangular subunits}

\subsection{Subunit design and structure classification}

We consider the assembly of cylindrical tubules from triangular subunits, as illustrated in figure~\ref{fig:fig1}a. The subunits are flat equilateral triangles, which bind edge to edge. Since a tubule is an object with non-zero curvature, that curvature must be encoded in some way into the subunit. We accomplish this goal by considering subunits that form a specified dihedral angle between their edges when they bind. Therefore, each triangle has associated with it three dihedral angles, one for each side. We consider a case where the interactions are specific and all three sides of the triangle are distinguishable. The simplest set of interactions that prescribe the assembly of a tubule is side 1 binds to side 1, side 2 to side 2, and side 3 to side 3. The specificity of these interactions is important in order to preserve the same local curvature everywhere across the assembly and to ensure that the pattern is deterministic. A schematic of such specific interactions and dihedral angles are shown in figure~\ref{fig:fig1}b. 

Any tubule formed from triangular subunits can be classified uniquely by a pair of indices. Consider a tubule as a rolled-up triangular lattice or a sheet with periodic boundary conditions along two parallel lines (figure~\ref{fig:fig1}c). Within this conceptual framework, a tubule can be constructed by choosing any two points on this plane and enforcing them to be periodic with one another. In other words, these two points overlap when the sheet is rolled up into a tubule, as in figure~\ref{fig:fig1}c.  To classify the tubule that forms, we count how many lattice edges need to be traversed between the periodic vertices. Taking the unit vectors of this tiling to be \textbf{m} and \textbf{n}, as shown in figure~\ref{fig:fig1}c, we can go between the vertices in \textit{m} and \textit{n} steps in the respective lattice directions, giving a total displacement of $\textbf{w} = m\textbf{m} + n\textbf{n}$. The corresponding tubule would be identified as $(m,n)$ and have a width of $w = \sqrt{m^2 + n^2 + mn}$. Here, the width $w$ refers to the width of the strip on the triangular lattice (figure~\ref{fig:fig1}c), which corresponds to the circumference of the closed tubule.

For any given tubule, there is a unique set of dihedral angles between adjacent triangular edges that will yield that desired structure. These values can be obtained by constraining the vertices of the lattice to lie on the surface of a cylinder and finding the angle between adjacent faces. Note that in this construction, the specified dihedral angles are constant along a given lattice direction, denoted by indices 1, 2, 3 in figure~\ref{fig:fig1}c. This constraint imposes an orientational order for the triangular subunits, which we will return to later.

\subsection{Computational methods}

We explore the assembly outcomes using grand canonical Monte Carlo simulations. Specifically, we use the event-driven Monte Carlo algorithm developed by Rotskoff et al.~\cite{rotskoff2018robust} and Li et al.~\cite{li2018large}, and adapted to tubules in Ref.~\cite{tyukodi2021thermodynamic}, in which an assembled structure exchanges subunits with a bath at fixed chemical potential. Each triangular subunit is modeled by three vertices and three straight edges connecting the vertices. The Hamiltonian of the system is given by,
\begin{eqnarray}
H = \frac{1}{2}\sum_\mathrm{Bound\ Edges} E_\mathrm{B} &+& \sum_\mathrm{Edges} \frac{1}{2}S (l - l_0)^2 \nonumber \\
&+& \sum_\mathrm{Adjacent\ Faces} \frac{1}{4} B(\theta - \theta_0)^2.
\end{eqnarray}
The first term is the binding energy. $E_\mathrm{B}$ is the energy difference between a pair of bound and unbound edges, and is set to be the same for all favorable interactions. The second term is the stretching energy. $S$ is the stretching modulus of the edge, $l$ is the instantaneous length of the edge, and $l_0$ is the stress-free length of the edge. The third term is the bending energy. $B$ is the bending modulus of the edge pairs and is again set to be the same for all edges. $\theta$ is the instantaneous dihedral angle between two subunits and $\theta_0$ is the preferred dihedral angle for a given tubule structure and type of edge-pair. See Supporting Information (SI) Section I for a detailed description of our computational methods. 

For a given set of input parameters---the bending modulus, the  lattice numbers \textit{m} and \textit{n} of the target, and the number of unique species---we perform one thousand independent simulations and analyze the distribution of tubule types that form. We prescribe the equilibrium dihedral angles, $\theta_0$, to favor (\textit{m}, 0) and enforce the binding rules specified previously. We tune the various energies in the Hamiltonian to keep the supersaturation low enough that the structure can nucleate, grow, and close near to equilibrium in a reasonable time scale. The simulation starts with a single subunit and grows by adding subunits onto the pre-existing structure, ending once the tubule has a length roughly three times its circumference. Finally, we determine the tubule type of the end state and compute the distribution of tubule structures for each condition. 

We only examine defect-free tubules in the following analysis. We consider a tubule to be defect-free if it has the same tubule structure, characterized by the same pair of indices (\emph{m},\emph{n}) along its entire length. While it is possible for defective tubules to form in our simulations, we find that the likelihood of forming a defective tubule does not depend on the number of subunit species used for assembly. However, the fraction of defect-free tubules does vary between 95\%-63\% depending on the targeted tubule width (see SI figure S3). We hypothesize that defect creation is related to the degree of supersaturation and the kinetics of growth near closure, rather than the nature of the free-energy landscape or the type of tubule that forms. This hypothesis is consistent with the type of defective tubules that we observe, which tend to have a central region that does not close properly, resulting in a tubule with a heterogeneous tubule type along its length. See SI Section II for a description of the algorithm that we use to determine the tubule type and for details on the defect rates.

\section{Results and discussion}
\subsection{The origin of off-target assemblies}

\begin{figure*}[tbh]
 \centering
 \includegraphics[width=\linewidth]{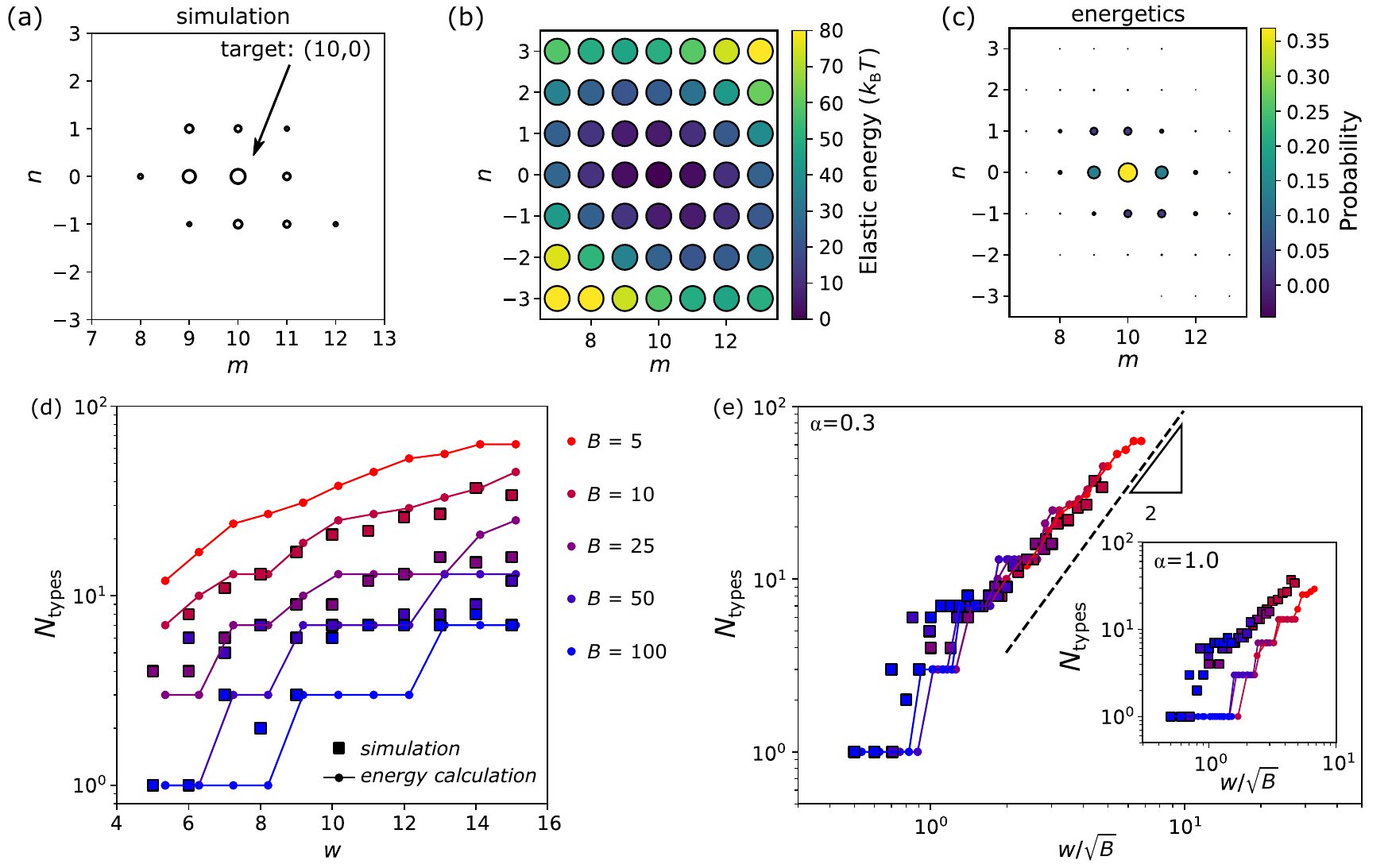}
 \caption{\textit{Reduction of target yield due to fluctuations}. (a) Distribution of tubules from a simulation with a target structure of (10,0) with $B=20$ \kbt/rad$^2$. The area of each circle is proportional to the number of defect-free structures that formed. (b) Theoretical elastic energy cost of forming different tubule types with dihedral angles that are designed to form a (10,0) tubule with $B = 20$ \kbt/rad$^2$. (c) Estimated probability for different assembled tubule types; circle size is proportional to the probability. The probability is estimated by $P(s)\sim e^{-\alpha E(s)/k_{\mathrm{B}}T}$, where $s$ is the state, $E(s)$ is the elastic energy of that state, and $\alpha$ is a factor adjusting the free-energy barrier described in the text.  (d) Number of tubule types $N_\textrm{types}$ that form for varying bending rigidity, $B$ (units of \kbt/rad$^2$), and target width, $w$. The points with connected lines are from the elastic energy calculations; the squares are the results from simulations. (e) Number of tubule types versus a rescaled width of $w/\sqrt{B}$. The dashed line is a guide to the eye and shows a power law with slope 2; the inset shows the same data, but with $\alpha=1.0$.}
 \label{fig:fig2}
\end{figure*}

The results of our simulations show that assembly yields a broad distribution of tubule types. Figure~\ref{fig:fig2}a shows an example of such a distribution. The bevel angles for this example were chosen to target a (10,0) structure. While we do see that the majority of tubules formed the target state, there is a broad distribution of adjacent states that have also formed, extending one to two lattice steps in both \textbf{m} and \textbf{n} directions. The probability for assembling off-target states falls off further from the target state as expected, since the dihedral angle differences, and hence the elastic energy cost, become more significant.

To understand the origin of this distribution, we consider the process by which assembly occurs and hypothesize that the distribution is dictated by the mechanism of tubule closure. Assembly begins when subunits come together to nucleate a curved sheet.  Once the extent of the sheet is large enough, it can close to form a short tubule, which then grows by the addition of subunits to its free ends. In the pre-closure state, thermal fluctuations can cause the sheet to close at larger or smaller widths around the target state, sampling the energies of different closed states.  We hypothesize that once the tubule closes, it is highly unlikely to open back into a sheet because doing so would require rupturing multiple edge-edge interactions simultaneously. Therefore, once the assembly has gone through closure and begun to extend, the tubule type is essentially fixed, even if it is not the global free-energy minimum. While the energy difference between off-target states and the target structure continues to increase as the tubule grows longer, without the ability to open again, the system can not equilibrate to the designed global energy minimum. Indeed, in simulations, we find that the tubule type does not change after a structure has closed (see figure S2c).

In addition to running simulations, we also estimate the distribution of structures that may arise during assembly from calculations of the elastic energy at closure. The penalty for forming off-target states comes from the elastic energy cost of not abiding by the prescribed dihedral angles. Recall that the bending energy, $E_\mathrm{\theta}$, for any binding site is given by $E_\mathrm{\theta}=\frac{1}{2} B(\theta-\theta_\mathrm{0})^2$. For the case of our triangular subunits, the elastic energy per subunit is half the sum over all three binding sites. When closure happens, it occurs for a finite number of subunits in the assembly, $\Nc$, all of which contribute to the elastic energy. The full elastic energy cost at closure is therefore
\begin{equation}
E_\mathrm{\theta}=\frac{1}{4}\Nc\sum_{i=1,2,3}B(\theta_i-\theta_{i,\mathrm{0}})^2,
\end{equation}
where the extra factor of $\frac{1}{2}$ comes from the fact that a single binding site is shared between two subunits. To estimate the closure size, we assume that subunit addition occurs isotropically, forming a circular disk with a diameter of the tubule width, $w$. Taking the ratio of the disk area to the area of a subunit gives $\Nc = \pi w^2/\sqrt{3}$. To get a more accurate accounting of the free-energy cost of forming misassembled structures, one would need to include the surface energies and the various entropic contributions, but we find that just considering the bulk elastic terms reproduces the scaling that we find in simulation.

We see the elastic energy at closure is lowest at the target state and increases by a few \kbt\ at the nearest off-target states, suggesting that these states are likely accessible in a system at finite temperature (figure~\ref{fig:fig2}b). To get an estimate of the tubule type distribution, we compute the probability of a structure $s$ according to $P(s)=e^{-\alpha E(s)/k_{\mathrm{B}}T}/Z$, where $Z$ is the partition function and $\alpha$ is a factor that adjusts the height of the free-energy barrier between the pre-closure and the closed states to account for missing contributions to the closure rate (e.g. deviations of pre-closed tubules from the assumed circular shape and other entropic effects). Supplementary simulations found a value of $\alpha=0.3$ and are described in SI Section III. Without this additional factor the energetics predict narrower distributions. Figure~\ref{fig:fig2}c shows the probabilities according to the energy landscape in figure~\ref{fig:fig2}b, which bears a similar shape and extent to the simulated distribution in figure~\ref{fig:fig2}a. 

Estimates of the number of accessible states follow power-law scalings with the tubule width and bending rigidity. From the energetics calculation, we make an estimate of the number of different structures, $N_\mathrm{types}$, by counting how many states have a probability greater than 0.25\%. From simulations, we look at the final tubule structures that form in four hundred different runs. We find that the number of different structures that assemble is sensitive to both the target width and the bending rigidity: The number of accessible states increases with increasing width and decreases with increasing bending rigidity (figure~\ref{fig:fig2}d). We find that these same data collapse to a single curve when rescaled by $w/\sqrt{B}$ (figure~\ref{fig:fig2}e). In the inset of figure~\ref{fig:fig2}e we show the same data but using $\alpha=1.0$ for the energetics, finding that the number of types predicted is lower than what we see in simulation. This implies that $\alpha$ is capturing some aspects of the kinetic processes during closure.

The scaling for the breadth of the tubule type distribution comes from a balance between the closure size and the fluctuations of the curvature of the sheet before closure. We consider the Helfrich energy of a curved sheet~\cite{helfrich1988intrinsic}, $E = \frac{1}{2}B A (\Delta \kappa)^2$, where $A$ is the area of the sheet and $\Delta \kappa$ is the deviation of its curvature away from the ideal curvature. We can approximate this curvature as $(\Delta w/w^2)$, where $\Delta w$ is the fluctuations of the width and $w^2$ is the area of the sheet. Looking at the size of these fluctuations on an energy scale of \kbt\ shows that $\Delta w \sim w/\sqrt{B}$. Therefore, the scaling that we find in figure~\ref{fig:fig2}e, $N_\mathrm{types}\sim \Delta w^2$, arises from the fact that thermal fluctuations populate a region of vertices around the target vertex with an area of $\Delta w^2$. 

These results illustrate a fundamental hurdle for self-limited assembly: in a thermal system, it is difficult to achieve specificity of a target state when the self-limited length scale is large compared to the subunit size. Small fluctuations of the dihedral angles between subunits become amplified as the number of subunits in the self-limited length scale increases. Even though the rigidity of individual dihedral angles may remain the same, the fluctuations of the self-limited length scale grow proportionally to the self-limited lengthscale itself, as we saw from the Helfrich energy. Compounding this effect, the process of irreversible closure prevents the assembly from visiting different states at later times to further relax. Therefore, the breadth of the distribution is driven by a kinetic process, which yields a larger variety of states than would be expected in equilibrium. It is this bottleneck to high-yield of the specific target that must be engineered around, either by making ever stiffer subunits, circumventing closure-control  by seeding nucleation with specific geometries, or by altering the energy landscape near closure. Here, we will explore this last direction by considering how multiple types of interacting subunits limit the accessible states at closure, and thereby prune off-target states from the distribution of final structures.

\subsection{Allowed tilings with multiple species}

\begin{figure}[tb]
\centering
\includegraphics[width=\linewidth]{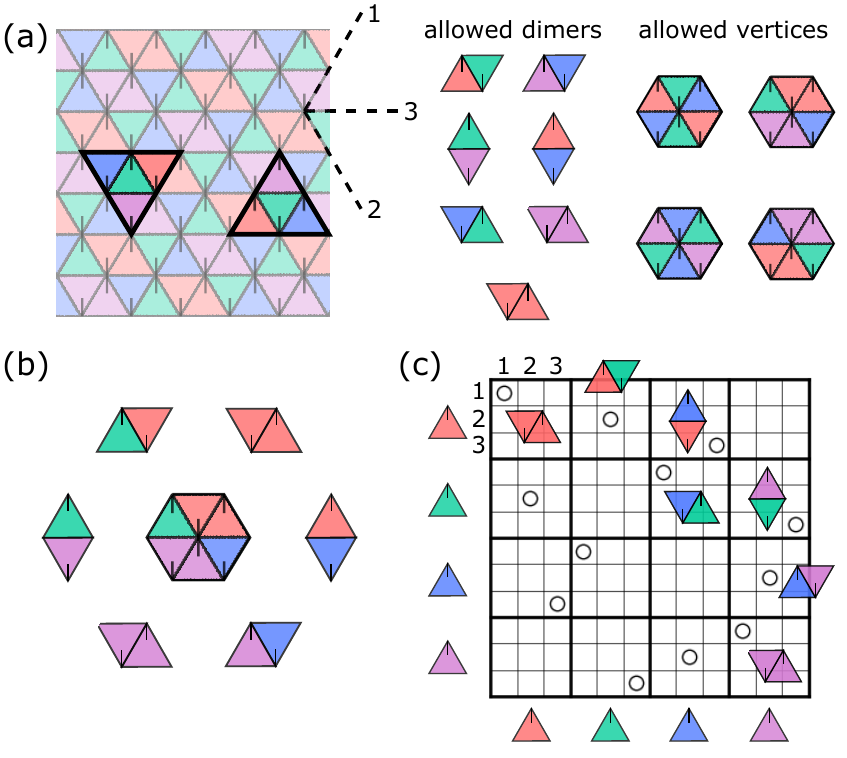}
\caption{
\textit{Rules for allowed tilings}. (a) An allowed four color tiling that can be wrapped into a tubule (pattern 4-4). Dashed lines labeled 1,2,3 show the three lattice directions that denote the specific sides of the triangles. Two green triangles and their neighbors are outlined to highlight their interactions. Tick marks on triangles are placed opposite of side 3 to illustrate the triangle orientation. Note that triangles may only appear in orientations with 180-degree rotations. Each specific interaction between two sides of triangles forms a unique dimer. To the right of the pattern are the only allowed dimers and the six-particle vertices that can be made from them. (b) For tilings, we require that each vertex be made up of six valid dimers, i.e. we do not allow non-bonded edges to appear in the tiling. (c) Each pattern can be represented by a symmetric interaction matrix. This matrix corresponds to the pattern shown in (a). Each circle shows a favorable interaction and has its associated dimer next to it. Colored triangles appear next to the columns and rows associated with each color of triangle. Each interior $3\times 3$ block shows the interactions for the three sides of one triangle species.}
\label{fig:fig3}
\end{figure}

To proceed, we extend our framework to allow for multiple species of triangles, where a species of triangle is a subunit with a distinct set of specific interactions encoded in its edges. The first task is to identify allowed patterns of multiple species that still satisfy the requirements imposed by the tubule geometry. This challenge involves finding periodic patterns in a triangular lattice with multiple colors of triangles. There are many ways that one can imagine periodically coloring a triangular lattice, but not all of these tilings will necessarily preserve the physical rules for tubule assembly. Here, we introduce three rules for multispecies tubule assembly. First, to have a deterministic assembly, there need to be unique interactions between subunits: Each side can only bind to one other side within the mixture of many subunit species. In the context of a tiling, this constraint means that once the colors of tiles adjacent to any subunit have been specified, there cannot be different neighbors at any other location in the tiling. Second, to impose constant dihedral angles along specific lattice directions, only rotations of 180 degrees for a given subunit type are allowed throughout the tiling (figure~\ref{fig:fig3}a). Third, if we view all interactions between subunits as creating allowed dimers with specific orientations, then we also must be able to construct closed vertices on the plane using either three or six of our specified interactions (figure~\ref{fig:fig3}b); this last constraint will force the system to assemble into a deterministic pattern that is the same everywhere in the tiling.

The concept of an interaction matrix helps to make these rules for subunit interactions more concrete. Elements of an interaction matrix will either be non-zero, prescribing an allowed attraction between specific sides of triangles, or zero, meaning no binding is allowed (figure~\ref{fig:fig3}c). In SI Section IV, we describe how the restrictions mentioned above can be translated to allowed constructions of interaction matrices. Following this construction, we enumerated all allowed tilings of a triangular sheet for up to ten different species of subunits. Figure~\ref{fig:fig3} shows one allowed tiling for four species, which are illustrated as four different colors. 

\begin{figure}[t]
 \centering
 \includegraphics[width=\linewidth]{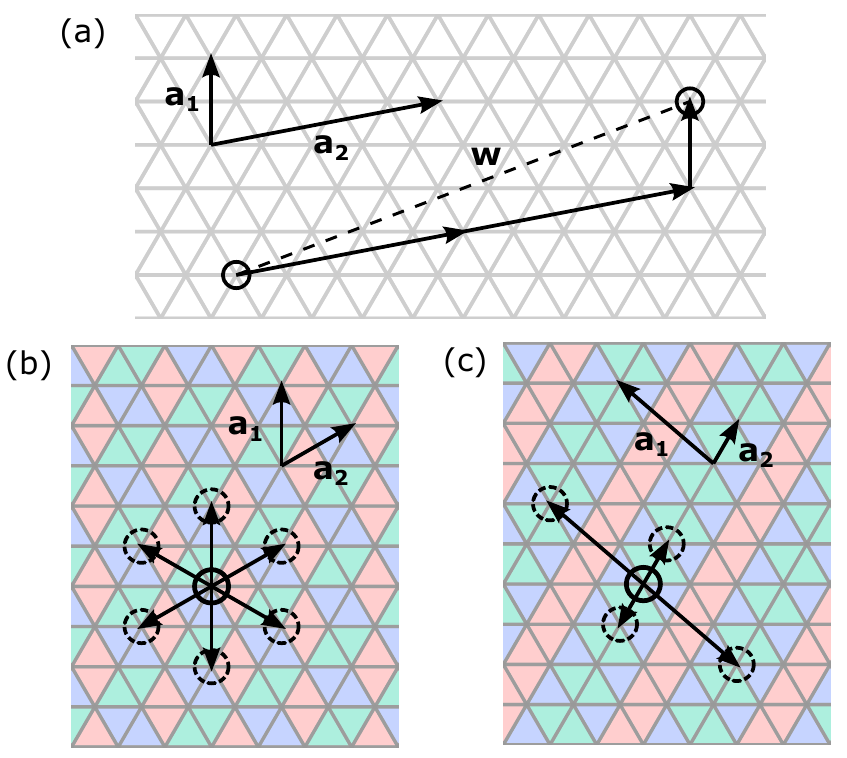}
 \caption{\textit{Restriction of allowed tubule states from multispecies patterns.} (a) Construction of a target tubule type with displacement vector, \textbf{w}, from a set of primitive vectors $\textbf{a}_1$ and $\textbf{a}_2$. The primitive vectors will differ depending on the specific tilings, but any allowed \textbf{w} must be a linear combination of primitive vectors. (b) and (c) show two different tilings for three species with different primitive vectors. The full circles show a potential target vertex for a tubule. The dashed circles show the nearest vertices accessible using only the primitive vectors of the tiling. (b) shows uniform distances between nearest vertices (pattern 3-1) while (c) has a large anisotropy between the $\textbf{a}_1$ and $\textbf{a}_2$ directions (pattern 3-0).
}
 \label{fig:fig6}
\end{figure}

Looking at the different tilings that can be wrapped into tubules, we notice that some of them will be more useful for restricting neighboring states than others. The most important feature of a tiling in this regard is its set of primitive vectors (figure~\ref{fig:fig4}). For any tiling we can define a pair of vectors, $\textbf{a}_1$ and $\textbf{a}_2$, that go between similar vertices in the tiling, have a minimal length, and are maximally orthogonal. These primitive vectors can then be used to identify which tubule types can be formed from a certain pattern. Recall that for tubule type $(m,n)$ there is an associated displacement vector $\textbf{w} = m\textbf{m} + n\textbf{n}$. If $\textbf{w}$ can be made from a linear combination of $\textbf{a}_1$ and $\textbf{a}_2$, then $(m,n)$ is an allowed state for that tiling (figure~\ref{fig:fig6}a). See the SI for patterns and interaction matrices.

When the two primitive vectors are large, the similar vertices are farther apart, leading to greater distances between allowed states. Furthermore, the closer in magnitude the primitive vectors are to one another, the more uniform the restriction of states will be around the target state. If there is a difference in the length of the primitive vectors then the tiling will be anisotropic and there will be an additional orientation dependence of the restricted states with respect to the orientation of the triangles. Two patterns made from three species of triangles are shown in figure~\ref{fig:fig6}b and c. From these two patterns we can see that figure~\ref{fig:fig6}b shows isotropic distances to nearby vertices while figure~\ref{fig:fig6}c shows anisotropic distances. Depending upon how \textbf{w} is aligned with respect to the anisotropic primitive vectors, the assembly outcome changes. For instance, if \textbf{w} is aligned along a shorter primitive vector, the density of states in that direction will be larger and will result in larger fluctuations of the widths of tubules that form. In the other case, when \textbf{w} is aligned along the longer primitive vector, there will be larger fluctuations in the chirality of tubules that form. To exemplify some of these points, we will look in detail at the patterns formed from two colors.

\subsection{Example of tilings with two colors}

\begin{figure}[t]
 \centering
 \includegraphics[width=\linewidth]{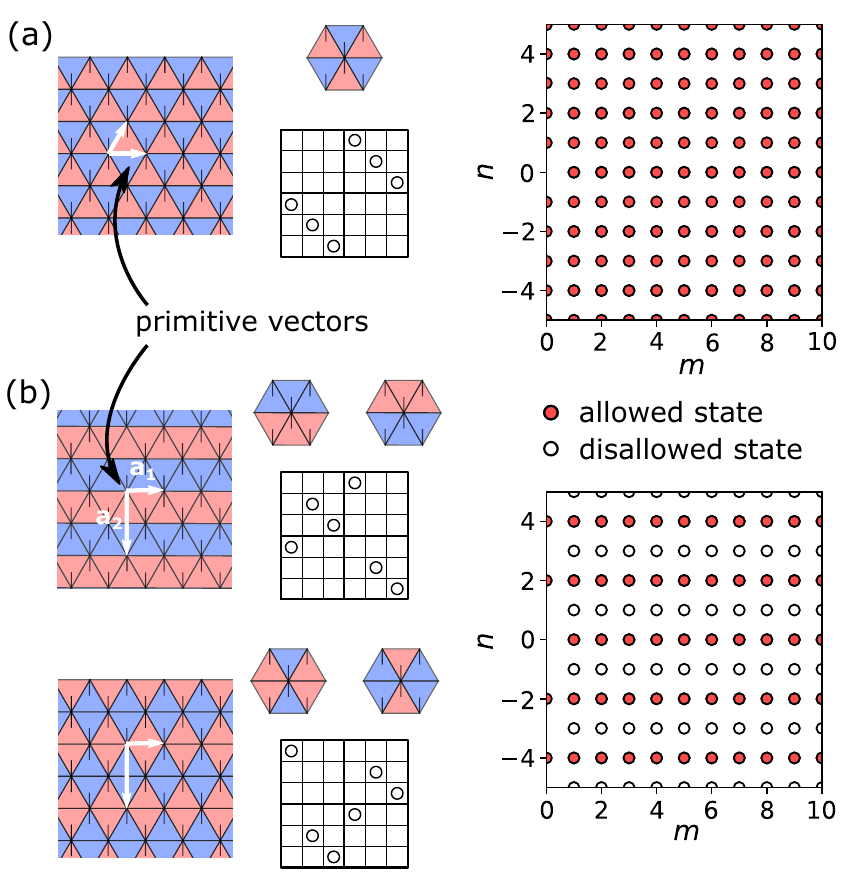}
 \caption{\textit{Two-color tilings and their allowed tubule types}. (a) A two-color tiling that results in no additional specificity (SI pattern 2-2). The arrows on the tiling denote the primitive vectors. For this pattern the single vertex is shown as well as its interaction matrix. On the right plot, filled red points show allowed (\textit{m,n}) tubules resulting from this tiling. (b) Two other two-color tilings with their respective unique vertices, interaction matrices, and primitive vectors (top: SI pattern 2-0; bottom: SI pattern 2-1). Note that the primitive vectors have the same lengths and orthogonality, resulting in the same set of allowed tubules, shown in the plot to the right. Open points show disallowed states.}
 \label{fig:fig4}
\end{figure}

To illustrate how multispecies tilings can limit the number of accessible off-target states, we first consider the allowed tilings composed of two species. We find that there are only three unique patterns that satisfy the restrictions for assembling tubules (figure~\ref{fig:fig4}), with each pattern having a unique interaction matrix required for assembly. For each pattern, we also compute the allowed tubule geometries. Interestingly, we find that one of the three patterns (figure~\ref{fig:fig4}a) does not give additional restrictions compared to a single-species tiling. In contrast, the other two tilings reduce the available states by half, albeit with the same restrictions as one another (figure~\ref{fig:fig4}b).

As mentioned above, we rationalize the observation that some tilings restrict the allowed states while others do not by looking at the primitive vectors of the tilings. For the pattern in figure~\ref{fig:fig4}a, we see that every vertex of the tiling is the same, meaning that all points in \textit{m,n}-space are allowed tubule types. This occurs because the length of its primitive vectors both match the subunit length. In contrast, the two patterns in figure~\ref{fig:fig4}b have two distinct vertices that cannot overlap with one another on a tubule. Because the two vertex types appear with equal frequency, each vertex can only bind with half of the total vertices in the pattern, thereby reducing the number of accessible states by a factor of two. Additionally, since the primitive vectors for the patterns in figure~\ref{fig:fig4}b are not equal length, the restrictions imposed on the allowed states are anisotropic. For this pattern, any value of \textit{m} is allowed, while only even values of \textit{n} are permitted. Depending on whether or not one wants to more tightly restrict the chirality of the structure or the available widths, one can adjust the orientation of the target displacement vector, \textbf{w}, with respect to the two primitive vectors.

A subtle point about the two patterns in figure~\ref{fig:fig4}b is that even though they produce the same restrictions on allowed states, they require different numbers of specific interactions to encode their patterns. For example, we see that there are five and four unique matrix elements for the top and bottom patterns in figure~\ref{fig:fig4}b, respectively. We hypothesize that reducing the number of interaction types to restrict a greater number of states will be an important design criteria for multiple-species experiments, since there will inevitably be a finite capacity for specific interactions beyond which undesired crosstalk between edges becomes non-negligible.

\subsection{Finding the minimal number of species required for high-yield assembly}

\begin{figure*}[t]
 \centering
 \includegraphics[width=\linewidth]{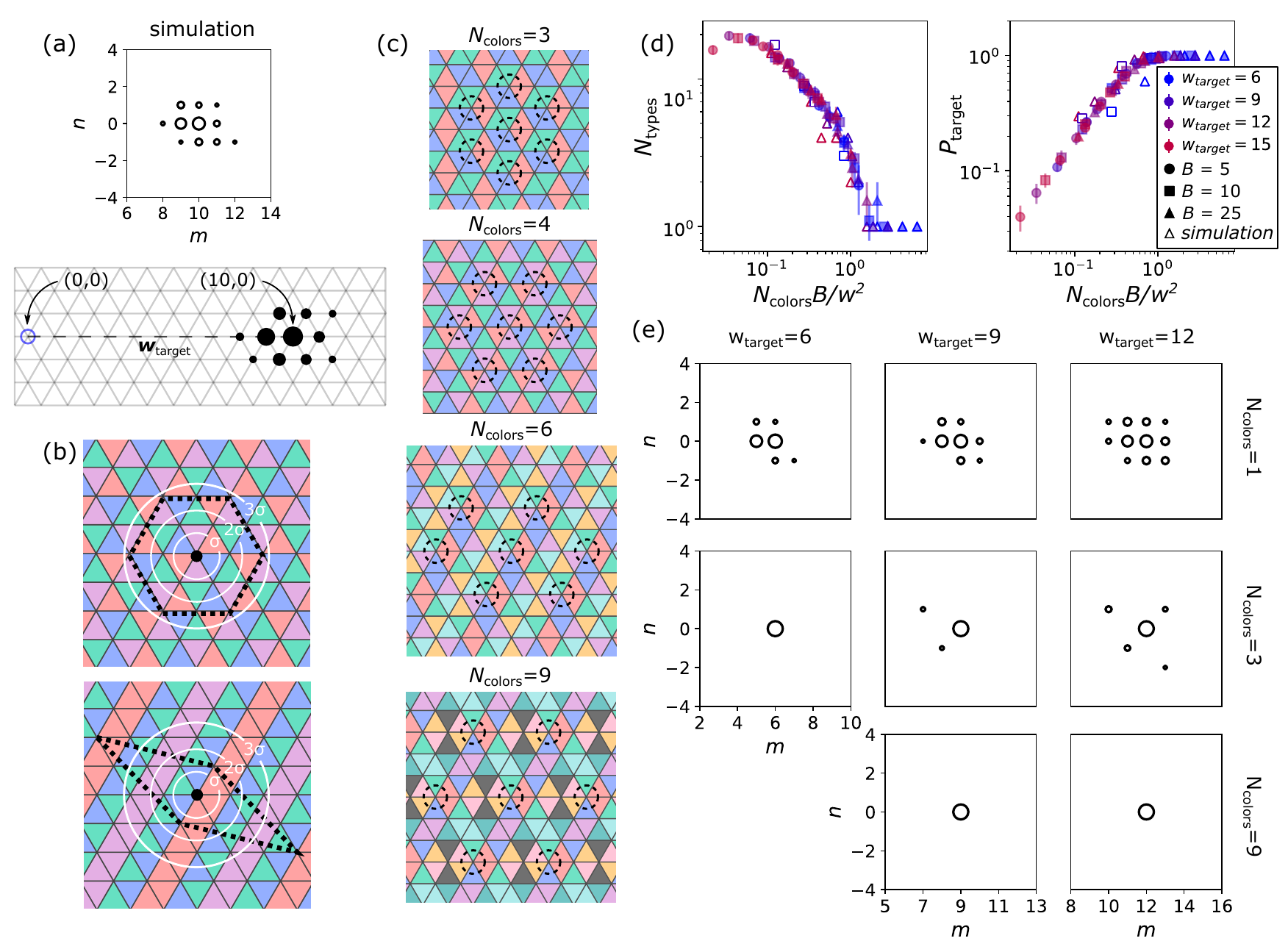}
 \caption{\textit{Increasing the target yield using multiple species}. (a) Tubule type distribution computed from simulations for a (10,0) target, represented in the \textit{m,n} space (top) and the same distribution overlaid on a triangular lattice (bottom). The diameter of each circles is proportional to the probability of observing that state. On the triangular lattice, we observe that the spatial distribution of states around the target vertex is roughly isotropic. (b) Examples of two four-color tilings (top: SI pattern 4-6; bottom: SI pattern 4-0). In both images the dotted line shows the region around a target vertex where no similar vertex lies. The white circles show the accessible area for closure sites of an assembly due to fluctuations of the pre-closure disk; higher standard deviation ($\sigma$) contours denote lower probability regions. The anisotropic tiling excludes fewer of the fluctuations. (c) Patterns used for calculating the reduction of assembly states as we increase the number of species, $N_\mathrm{colors}$(from top to bottom SI patterns are 3-1, 4-6, 6-3, 9-2). All but the $N_\mathrm{colors}=6$ pattern are considered isotropic patterns. (d) Number of accessible tubule types, $N_\mathrm{types}$, and probability of forming the target state, $P_\mathrm{target}$, versus the rescaled number of colors, $N_\mathrm{colors}B/w^2$. Results from energetics calculations are the solid points; results from simulations are the open points. We observe a nearly perfect yield of the target state when $N_\mathrm{colors}B/w^2 \gtrsim 1$. (e) Examples of simulated tubule distributions for target states (6,0), (9,0), and (12,0) with $N_\mathrm{colors}=1,3,9$. Larger target structures have broader initial distributions, but all exhibit a high probability of the target state with a small increase in the number of species for the tiling.
}
 \label{fig:fig5}
\end{figure*}

Now that we understand how a tiling with multiple species can change the number of accessible states, we explore how to design a system that targets a single state with a high yield. As we saw in figure~\ref{fig:fig2}d, the number of off-target states, $N_\mathrm{types}$, increases as the area of fluctuations around the target vertex, $\Delta w^2$, increases. To compensate for this effect, we expect that the required primitive vector length of a tiling should be comparable to the fluctuations of the self-limited length scale to achieve high yield of the target state. Figure~\ref{fig:fig5}a shows how a distribution of tubule types relates to the spatial distribution of allowed states around the target vertex. We see that the accessible vertices are nearly uniformly distributed around the target vertex, with a slightly larger variation in the \textbf{w} direction. Therefore, we expect that tilings that have primitive vectors of similar lengths will eliminate off-target states most effectively. Figure~\ref{fig:fig5}b shows examples of two tilings and their respective excluded regions. The first pattern shows an isotropic tiling with the nearest accessible vertices shown connected by the dotted line. The second pattern shows an extreme case in which the excluded region is highly anisotropic. In the short direction, smaller fluctuations would be needed to access another allowed vertex. Going forward, we restrict ourselves to the use of isotropic or near-isotropic tilings; the patterns used for the different numbers of species are shown in figure~\ref{fig:fig5}c.

To see how increasing the number of species, $N_\mathrm{colors}$ affects the assembly specificity, we perform both simulations and energy calculations as before. Specifically, we calculate the number of accessible states for tubules of different target widths, $w_\mathrm{target}$, and bending rigidity, which are assembled from different numbers of subunit species. 

We find that the assembly yield of the target state increases as the number of species increases. Figure~\ref{fig:fig5}d shows the number of tubule types that form, $N_\mathrm{types}$ (left), as well as the probability of forming the target state, $P_\mathrm{target}$ (right), as we change the number of species, rescaled by $w^2/B$. Recalling that the number of tubule types for a single species, $N_\mathrm{types}$, grows as $w^2/B$ (figure~\ref{fig:fig2}e), we consider the quantity $w^2/B$ as the area of the fluctuations around the target vertex. For isotropic tilings the primitive vectors have a length that grows as $\sqrt{N_\mathrm{colors}}$, meaning the excluded area around a target vertex will grow linearly with the number of species. When $N_\mathrm{colors}B/w^2\approx 1$  the area of the disallowed region and the area of fluctuations become comparable. At this point we observe full specificity in both $N_\mathrm{types}$ and $P_\mathrm{target}$. Both the energetics and the simulation show the same scaling. 

To clearly illustrate how adding additional species impacts the tubule distributions, we show examples of three different target widths for $N_\mathrm{colors}=1,3,9$ in figure~\ref{fig:fig5}e. As the number of species is increased there is a reduction in the number of types of structures that form, with a corresponding increase in the fraction that form the target state. Note that even though the specificity of the target is increased, the extent of the off-target states is not impacted, as seen most clearly in the three-species case. These results further illustrate that the minimum number of subunit species to achieve full specificity is proportional to the size of the fluctuations of the system.

\section{Conclusions}

In this report, we have shown that multiple species with specific interactions can be used to reduce the assembly of off-target structures for cylindrical tubules. As more species are added to the tiling, the distance between similar vertices increases, corresponding to an area of disallowed states around that target vertex. Full specificity of the target can be achieved when the area of the disallowed region encompasses all off-target states that would have been accessible due to thermal fluctuations. In using multiple species, there is a trade-off between the increase in complexity of the system and the benefit of greater specificity. Therefore, design rules, like the one we develop here, are essential to program the assembly of self-limiting structures in as economical a way as possible. An important note is that while this strategy successfully reduces nearby states for specific targets, the use of multiple species also limits the available target states that can be designed. This idea for how to engineer the free-energy landscape near a target structure should hold for other types of self-limited assembly as well.

An aspect of assembly that has not been discussed in this work is the dynamics of assembly. As the number of subunit species increases, we expect the time for nucleation and growth to increase significantly. This slowing-down of the dynamics of assembly places further emphasis on the need to design the number of unique subunits in as economical a way as possible. Beyond economical design, other strategies might also be combined with multispecies assembly to help overcome kinetic bottlenecks. Nature offers one strategy to improve the target yield without dramatically slowing the assembly kinetics: seeded nucleation. For example, \textit{in vivo}, microtubules---cylindrical cytoskeletal filaments formed from proteins---assemble with a narrow distribution of diameters with the help of seeds~\cite{roostalu2017microtubule}. Without such a mechanism to control the initial diameter of the filaments, an array of different tubule structures with varying width and chirality can be seen~\cite{chretien1991new,sui2010structural}. Having seeds allows the system to target a specific structure without sacrificing the kinetics. However, this approach would offer its own set of challenges for synthetic self-assembly with respect to the creation and purity of templates that are used to seed nucleation.

Aside from controlling the assembly specificity, there are other interesting directions that can be explored with the tilings that we have identified. Foremost is the possibility for increasing the addressability of an assembled structure with a tunable length-scale that is not limited to the particle size or the self-limited length scale. In the case of subunits made from DNA origami, we can imagine creating unique addressable sites for conjugating molecules or other small particles to specific subunits within a completx triangular lattice~\cite{kuzyk2018dna,dey2021dna,sigl2021programmable}. This strategy could be a way of patterning structures with receptors with certain biological functions or with nanoparticles to create materials with unique photonic responses. Similarly, by leaving out certain species, one could create a user-prescribed pattern of holes in the final structure, enabling the assembly of structures with tunable porosity.

Lastly, we have only explored the role of multiple species in the context of geometrically identical subunits with specific interactions. In terms of self-limited assemblies, there are other closed structures that have variable curvature throughout, such as toroids or helicoids. By engineering a set of allowed interactions between multiple subunits geometries, we could envision changing curvature or edge length for each component as well. Specifically, the linear-tilings that we have found (see SI figure S11 pattern 10-4 as an example) could be used to construct more complex manifolds.

\begin{acknowledgments}
We thank Seth Fraden and Gregory M. Grason for sparking an interest in this problem. T.E.V, H.F., D.H., M.F.H., and W.B.R. designed the research, T.E.V preformed the energetics calculations, T.E.V. and D.H. performed the pattern search, H.F. performed the simulations, B.T. implemented the simulation algorithm, and all authors participated in writing the manuscript. This work is supported by the Brandeis University Materials Research Science and Engineering Center, which is funded by the National Science Foundation under award number DMR-2011846, the National Institute of General Medical Sciences R01GM108021, the NSF XSEDE computing resources allocation TG-MCB090163 and the Brandeis HPCC, which is partially supported by the NSF through DMR-MRSEC 2011846 and OAC-1920147.
\end{acknowledgments}

\bibliography{main.bib}

\end{document}


\title{Supplemental Information for ``Tiling a tubule: How increasing complexity improves the yield of self-limited assembly"}

\author{Thomas E. Videb\ae k, Huang Fang, Daichi Hayakawa, Botond Tyukodi, Michael F. Hagan, W. Benjamin Rogers}
\affiliation{Martin A Fisher School of Physics, Brandeis University, Waltham, MA, 02453, USA}

\maketitle

\setcounter{figure}{0}
\makeatletter 
\renewcommand{\thefigure}{S\arabic{figure}}



\section{Implementation of the Monte Carlo algorithm}

We adapt the Monte Carlo algorithm developed by Rotskoff et al. and Li et al.~\cite{rotskoff2018robust,li2018large} and a data structure based on OpenMesh~\cite{bischoff2002openmesh}. 
Each triangular monomer is modeled as a triangular mesh element, which is composed of three vertices and three straight edges connecting the vertices. The Hamiltonian of the system is given by,
\begin{gather*}
 H = \frac{1}{2}\sum_\mathrm{Bound\ Edges} E_\mathrm{B} + \sum_\mathrm{Edges} \frac{1}{2}S (l - l_0)^2 + \sum_\mathrm{Adjacent\ Faces} \frac{1}{4} B(\theta - \theta_0)^2.
\end{gather*}
where $E_\mathrm{B}$ is the binding energy, $S$ is the stretching modulus, $l$ is the instantaneous length of the edge, $l_0$ is the stress-free length of the edge, $B$ is the bending modulus of the edge, $\theta$ is the instantaneous dihedral angle between two monomers that bind through an edge, and $\theta_0$ is the preferred dihedral angle that favors a certain tubule structure.

\begin{figure}[tb]
\centering
\includegraphics[width =0.5 \linewidth]{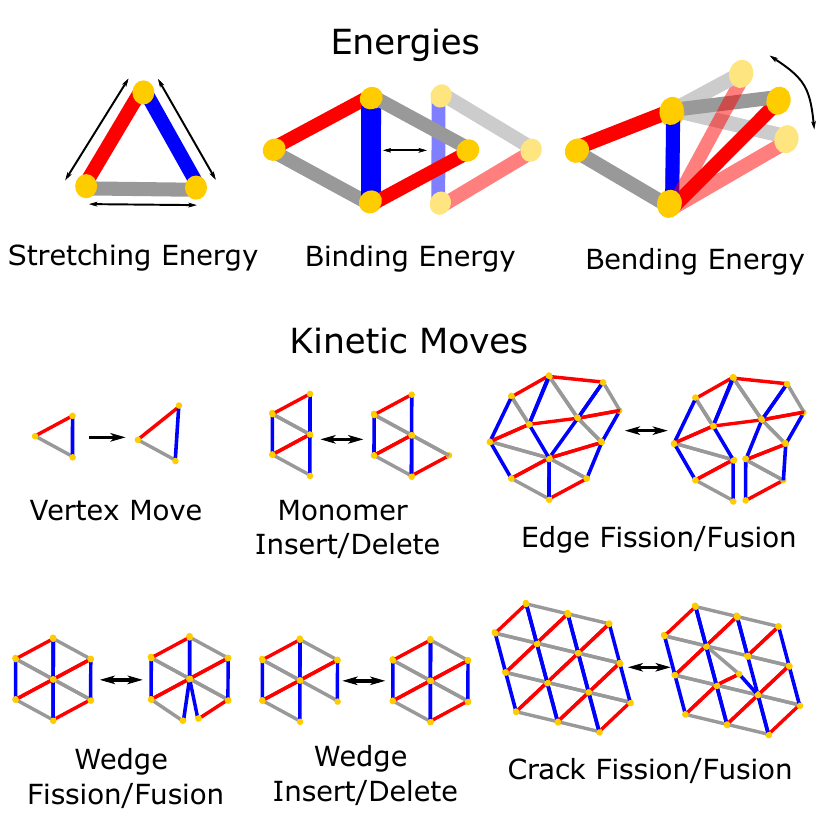}
\caption{\textit{Energies and kinetic moves in the simulation}. Energies include the stretching energy, the binding energy, and the bending energy. Kinetic moves include a vertex move, a monomer insertion/deletion move, an edge fission/fusion move, a wedge fission/fusion move, a wedge insertion/deletion move, and a crack fission/fusion move, as illustrated.}
\label{FigS6.Simulation}
\end{figure}
 
With this Hamiltonian, the system evolves under eleven different kinetic moves, as shown in figure~\ref{FigS6.Simulation}. The vertex move is attempted the most frequently. When attempting a vertex move, all the vertices that share the same position move collectively to a new position under a uniform distribution, $U$:
\begin{gather*}
    x \leftarrow x+U(-\dMax,\dMax) \\
    y \leftarrow y+U(-\dMax,\dMax) \\
    z \leftarrow z+U(-\dMax,\dMax),
\end{gather*}
where $\dMax$ is the maximum displacement, which is adjusted between $0.01l_0$ to $0.1l_0$ to optimize convergence to the equilibrium. $l_0$ is the stress-free length of the edge. The acceptance probability of a vertex move attempt is $\textrm{min}[1,e^{-\frac{\Delta E}{k_\mathrm{B}T}}]$, where $\Delta E$ is the energy difference between the structure before and after the move. The other kinetic moves are attempted less frequently than the vertex move. Monomer insertion/deletion moves and edge fission/fusion moves are attempted 1,000 times less frequently, while wedge fission/fusion moves, wedge insertion/deletion moves, and crack fission/fusion moves are attempted 10,000 times less frequently. We select the relative frequencies of the various moves to yield the assembly of a majority of defect-free tubules for the binding energies and bending rigidity that we choose. While the acceptance probabilities are different for different types of moves, they all guarantee detailed balance (for details, see the supplementary information of Ref.~\cite{tyukodi2021thermodynamic}). Furthermore, different values of the relative frequencies result in minimal changes to the resulting tubule distribution. We demonstrate this result by running simulations with different attempt frequencies of the edge fusion/fission moves and measuring the tubule type distribution. The mean of the distribution increases by roughly 5\% as the attempt frequency of edge fusion/fission moves decreases by two orders of magnitude, while the breadth of the distribution changes by less than 3$\%$.


The input parameters for simulations are set in the following ranges.
We set the binding energy of favorable edge-edge interactions to between $-5.2$--$-6.4$~\kbt and the binding energy of non-complementary interactions to 1000 \kbt. We set the bending rigidity to between 10--100~\kbt/rad$^2$. All simulations are run with the same stretching modulus of 200~\kbt\ and the chemical potential is held at -3~\kbt\ for each species. We choose this parameter space to keep the supersaturation low so that the structure can nucleate, grow, close, and approach steady-state tubule growth in a reasonable time scale.

\section{Identification of tubule structure}

\begin{figure}[tbh]
\centering
\includegraphics[width = 0.8\linewidth]{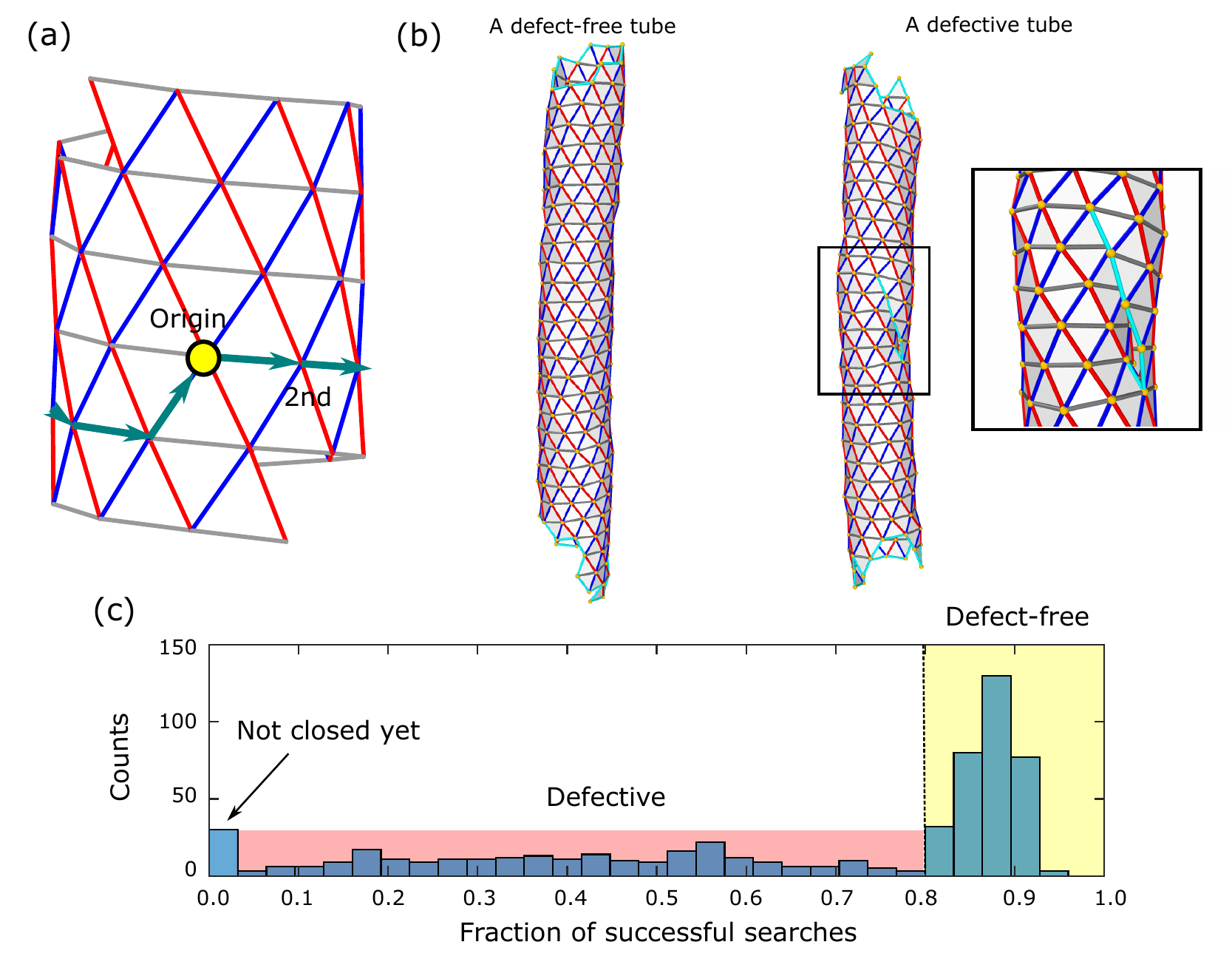}
\caption{\textit{Algorithm for identifying tubule types and determining their distributions.} (a) illustrates our search algorithm for determining the tubule type. The yellow circle shows the origin, while the solid turquoise arrows show the shortest pathway. (b) shows examples of defect-free and defective tubules of tubule structure. Only defect-free tubules are counted in the statistics for the tubule type distribution in the main text. In the snapshots, free edges are labeled in cyan. (c) shows the distribution of the fraction of successful searches for 500 simulations run with the same set of parameters. We perform the search algorithm for all of the vertices in the final structure and compute the fraction of successful searches for each simulation. We identify a structure as a defect-free tubule only when over 80\% of the searches are successful. Simulation parameters: $E_\mathrm{B}=-6.4~\ktMM,\ B=25~\ktMM/\mathrm{rad}^2,\ m=15,\ k=3$}
\label{FigS7.TubeID}
\end{figure}

We classify the tubules that form in simulations by searching for the shortest path around the circumference of the structure, moving along the edges of the triangles. Figure~\ref{FigS7.TubeID}a illustrates one such path. We choose a random vertex as the origin. Then, we search for the next vertex connected to the origin by a gray edge, which we call the second vertex. The vector pointing from the origin to the second vertex defines the direction of the search. Then, we search for vertices along either the red or the blue edges and measure the distance between those vertices and the origin. If those vertices are farther from the origin, we move to the next vertex that is connected through a gray edge. If those vertices are closer to the origin, we move along the red/blue edge until we reach the boundary or return to the origin. If the search reaches the boundary, we move to the next vertex that is connected through a gray edge.  
This search continues until we return to the origin or terminates if the last vertex connected through a gray edge is on the boundary. If the search can return to the origin, we label it as a success, otherwise we label it as a failure. The lattice numbers $m$ and $n$ are determined by the number of edges traversed on this pathway. $m$ is the number of the gray edges, while $|n|$ equals to the number of red or blue edges. $n$ is positive if the pathway follows blue edges and negative if the pathway follows red edges.

To identify defective tubules and to prevent incorrect identification, we perform the search algorithm described above for every vertex in the final assembly and measure the successful search rate. Figure~\ref{FigS7.TubeID}c shows a distribution of the fraction of successful searches of 500 simulations under the same parameter set. By comparing the simulation snapshots and the distribution, we find the tubes are mostly defect-free around the peak at 0.9 and mostly defective when the fraction of successful searches is less than 0.8. Thus, we identify a tubule structure as defect-free if over 80\% of the searches find the same tubule type.

\begin{figure}[hbt]
\centering
\includegraphics[width = 0.9\linewidth]{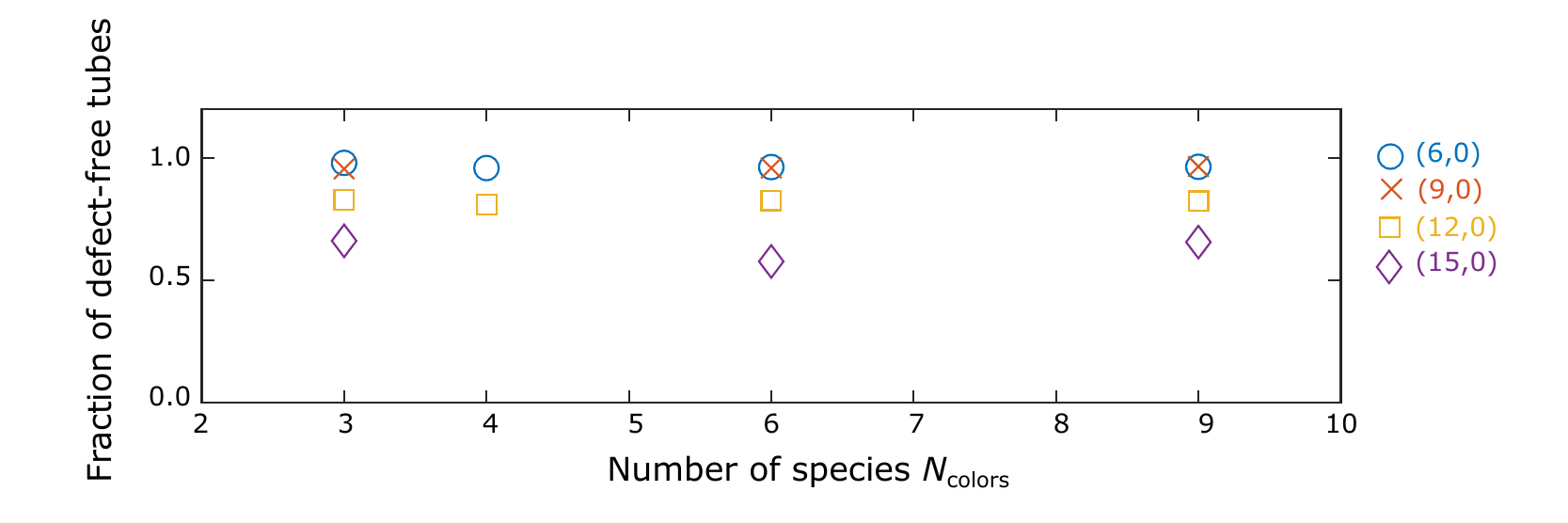}
\caption{\textit{Fraction of defect-free tubes} We compute the fraction of defect-free tubules for different numbers of subunit species. Different colors represent different target structures. Simulation parameters: $E_\mathrm{B}=-6.4~\ktMM,\ B=25~\ktMM/\mathrm{rad}^2$ }
\label{FigS10.DefectiveRate}
\end{figure}

The fraction of defect-free tubules is generally greater than 60\% for each parameter set we explore. As figure~\ref{FigS10.DefectiveRate} shows, we find the fraction of defect-free tubes changes with the size of the target structure but does not change significantly with the number of species we use for assembly. Due to the varying number of defect-free tubes for each distribution, all results in the main text are shown for a subset of 400 randomly chosen defect-free tubules, so that all distributions are sampled to the same extent.

\begin{figure}[tbh]
\centering
\includegraphics[width = 0.4\linewidth]{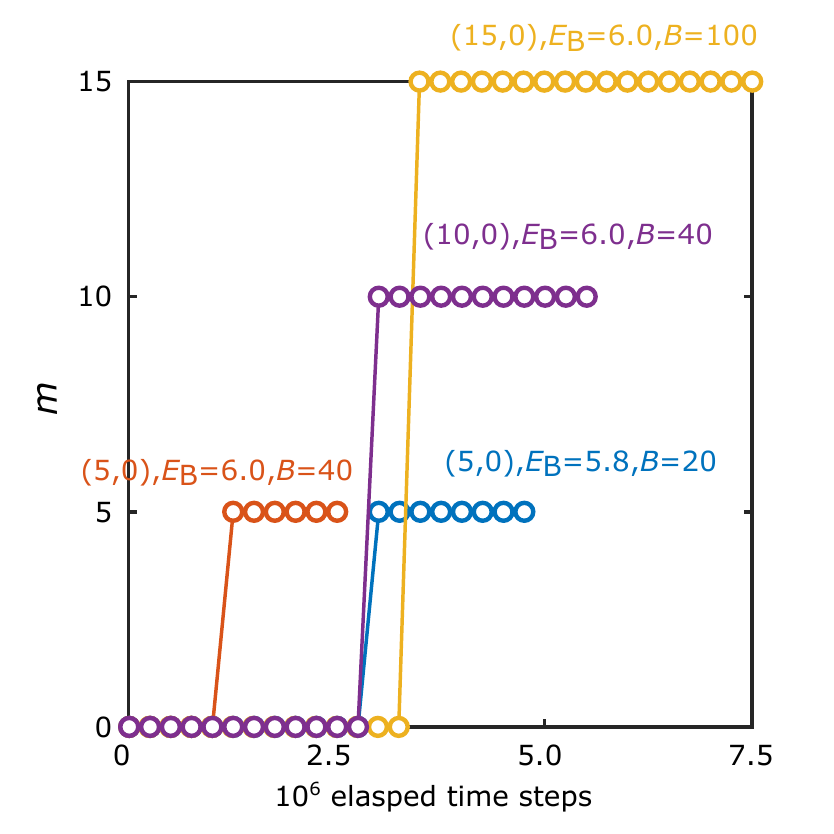}
\caption{\textit{Time independence of tubule type identification.} 
We show examples for four tubule structures and their identified \textit{m}-index during a simulation. When the structure is open, the \emph{m}-index equals to 0. At closure, the tubule can be identified and does not change for the remainder of the run. All examples shown here are for \textit{n}=0.}
\label{FigS9.TubeTypevsTime}
\end{figure}

Performing the tubule-type identification at different time points along the simulation trajectory shows that the tubule type is fixed after closure. Figure~\ref{FigS9.TubeTypevsTime} shows examples of the tubule type for a selection of simulation parameters over time. We see that once closure occurs, the tubule type is fixed and does not vary over the rest of the simulation. This observation is consistent over the full range of parameters that we explore.

\section{Evaluating the free-energy barrier for closure}

\begin{figure}[tb]
\centering
\includegraphics[width = \linewidth]{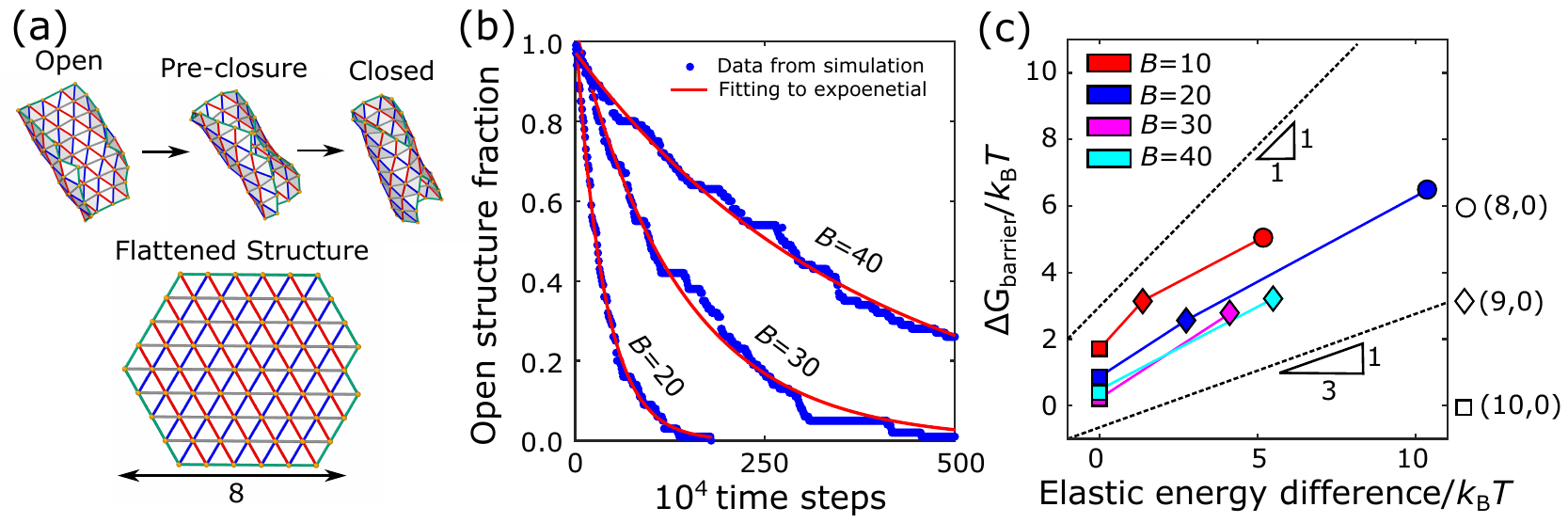}
\caption{\textit{Evaluating the free-energy barrier to closure}. (a) shows an example of a system that we simulate. The flattened structure is hexagonal, with a width of 8 edges across. The favored tubule type is (10,0), meaning that there will be an energy cost for closure due to the curvature difference between the closed and open configurations at this size. The trajectory shows three snapshots corresponding to the open state, pre-closure state, and closed state. (b) Measurements of the fraction of open structures as a function of time. The blue points are based on the statistics collected from the simulation; the red lines are exponential fits. (c) A comparison between the free-energy barrier heights and the estimated elastic energy differences between the pre-closure configurations and the stress-free configurations. Different colors correspond to different values of the bending modulus $B$ and different symbol shapes correspond to closure into different tubule types. }
\label{FigS8.Energybarrier}
\end{figure}

We evaluate the actual free-energy barrier to closure by measuring the closure rate and comparing it to the estimate based on the elastic energy difference alone. Figure~\ref{FigS8.Energybarrier}a shows the system that we investigated. When flattened onto a two-dimensional space, it has the form of a hexagonal sheet. We choose this geometry because the binding energy is the same for all three edges and the structure grows isotropically before closure. To measure the closure rate, we allow vertex moves only and sample 200 open structures as they relax to  equilibrium  for the same parameter set. Next, we allow both vertex moves and edge fission/fusion moves for the same parameter set and measure the fraction of the structures that remain open as a function of time. Figure~\ref{FigS8.Energybarrier}b shows that the fraction of open structures decays as a function of  time. By fitting a single exponential to this data, we extract the closure rate $k_\mathrm{closure}=1/\tau$ from the characteristic time $\tau$ of the exponential decay. This rate depends on the bending modulus. We expect the closure rate to be related to the free-energy barrier for closure, $\Delta G_\mathrm{barrier}$, by the following relationship,
\begin{gather*}
 k_\mathrm{closure} = k_0 \exp \left( -\frac{\Delta G_\mathrm{barrier}}{\ktMM} \right)
\end{gather*}
where $k_0$ is a kinetic factor. As the bending modulus $B$ increases, the free-energy barrier for closure into off-target states also increases due to the increasing elastic cost of unfavorable curvature, which causes the closure rate to decrease.

The free-energy barrier that we infer from simulation results is systematically lower than the estimate based on the elastic energy alone. Figure~\ref{FigS8.Energybarrier}c compares the measured $\Delta G_\mathrm{barrier}$ to the estimate of the elastic energy difference between the pre-closure configuration and the stress-free open configuration. We compute the elastic energy difference by assuming that all the monomers in the pre-closure configuration have the same dihedral angle as the monomers in the closed configuration, and that the dihedral angle in the stress-free configuration is the same as the equilibrium dihedral angle. For the same bending modulus $B$, the free-energy barrier $\Delta G_\mathrm{barrier}$ is roughly $\sim \frac{1}{3}$ of the estimated elastic energy cost. We hypothesize that this factor arises because the closure happens locally and thermal fluctuations need not bend the entire sheet to induce closure. In addition, our estimate does not account for the effect of thermal fluctuations on the stress-free configuration or other entropic effects. Therefore, we define a factor $\alpha < 1$, which relates the estimated elastic energy cost to the measured free-energy barrier height. For tubules smaller than (15,0), we see that $\alpha$ is roughly constant at $\alpha=0.3$. We use this value for our analysis presented in the main text.  

\section{Search for color-tiling patterns}

\begin{figure*}[tbh]
    \centering
    \includegraphics[width=1\linewidth]{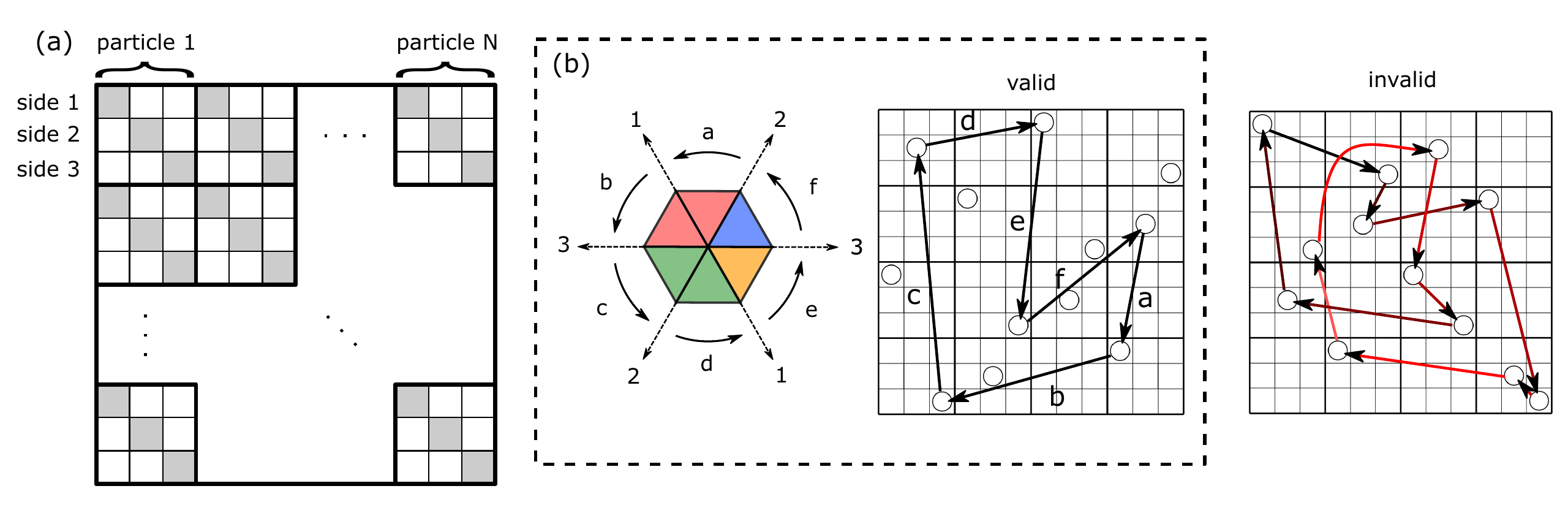}
    \caption{\textit{Structure of interaction matrices}. (a) Schematic of the arrangement of an interaction matrix. Each group of three rows and columns constitutes a sub-matrix showing the specific side edge-to-edge interactions of particles. The darkened boxes show the only elements that are allowed to be non-zero, as described in the text. (b) The box contains an example of a vertex loop in an interaction matrix. The left panel shows the order of jumps between particle interactions, and the right panel shows the corresponding loop. The panel outside of the box shows another interaction matrix that contains a loop that does not create a valid tiling.}
    \label{fig:figS1}
\end{figure*}

To search for different tiling patterns we make use of the fact that we can represent the interactions between different sub-units using an interaction matrix. Typically, for particles that have isotropic interactions, an interaction matrix, $\mathbf{I}$, is symmetric with elements such that if $\mathbf{I}_{ij}=\mathbf{I}_{ji}=1$ then particles types \textit{i} and \textit{j} can bind, and if $\mathbf{I}_{ij}=0$ then the particles cannot bind. For the case of our triangular sub-units we need to be able to account for the valency of the interactions to keep track of which side of each particle binds to which side of another. If we have $N$ particles with $S$ sides then the $\textbf{I}$ will have $N\times S$ columns and rows. We use the convention that for a row or column \textit{i} the side of the interaction is identified by $\mathrm{mod}(\textit{i}, 3) = s$ (\textit{s} being the index of a side) and the particle index is $\mathrm{floor}(\textit{i}/3)=n$ (\textit{n} being the index of a particle). This arranges $\mathbf{I}$ into \textit{N} sub-matrices of rank \textit{S} where each sub-matrix shows the side interactions between two species of particles. Figure~\ref{fig:figS1}a illustrates the  layout of such a matrix.

As mentioned in the main text we have three restrictions that we impose on the particle interactions to ensure tubule formation. The first two are: (i) interactions between particles are unique (as well as having all sides of all particles interact) and (ii) interactions must occur on the same side of a particle, i.e. only allowing rotations of 180 degrees of particles with respect to the lattice. These two restrictions can be translated into restrictions on the form of interaction matrices we can have. Restriction (i) means that all rows (and hence columns) of $\mathbf{I}$ must have one and only one non-zero element; this ensures that any tiling we generate will be deterministic. Restriction (ii) means that side \textit{s} of any particle can only interact with side \textit{s} of another particle, i.e. each sub-matrix can only have diagonal elements be non-zero.

With these restrictions in place on our interaction matrices, we now enumerate all possible interaction matrices that satisfy these rules. Then, we discard all matrices that are the same up to exchanges of particle indices (recoloring the patterns) or exchanges of side index (rotations or reflections of the patterns). We also discard interaction matrices that form patterns that do not involve all the particles; i.e. there is a subset of particles whose interactions make up a valid interaction matrix that produces a deterministic tiling pattern. An easy way to see that these exist is that we can take any valid matrix and append it to the lower corner of another matrix.

At this point we find that our two restrictions are not sufficient to generate deterministic patterns. We find that some interaction matrices allow for the collection of particles around a vertex on the lattice to not be fully specified. To address this, we note that there exist interaction loops within these matrices that correspond to a loop around a vertex. Consider a vertex of particles, such as in figure~\ref{fig:figS1}b. Starting with particle 1, we can ask which particle is clockwise of this one; in this case it is particle 2 and the interaction is occurring on side 1. We can continue this process until we return to our original particle. Following the six interactions that it took to bring us around a vertex, we can construct a rule for how to walk along this loop in the interaction matrix:
\begin{quote}
If the current element is in sub-matrix (\textit{n,m}) with side \textit{s}, then the next element in the vertex loop has the first index of the sub-matrix of \textit{m} with side mod(\textit{s-1}, 3); there can only be one such element.
\end{quote}
Noting that we allow for 180 degree rotations, these vertex loops can only be length three or six in a valid pattern. Examples of a valid and an invalid interaction matrix are shown  in figure~\ref{fig:figS1}b. Due to this possibility, we discard interaction matrices that have vertex loops with incorrect length. This corresponds to the third restriction in the text --- that all vertices be made up of valid dimer interactions.

\begin{figure*}[tbh]
    \centering
    \includegraphics[width=1\linewidth]{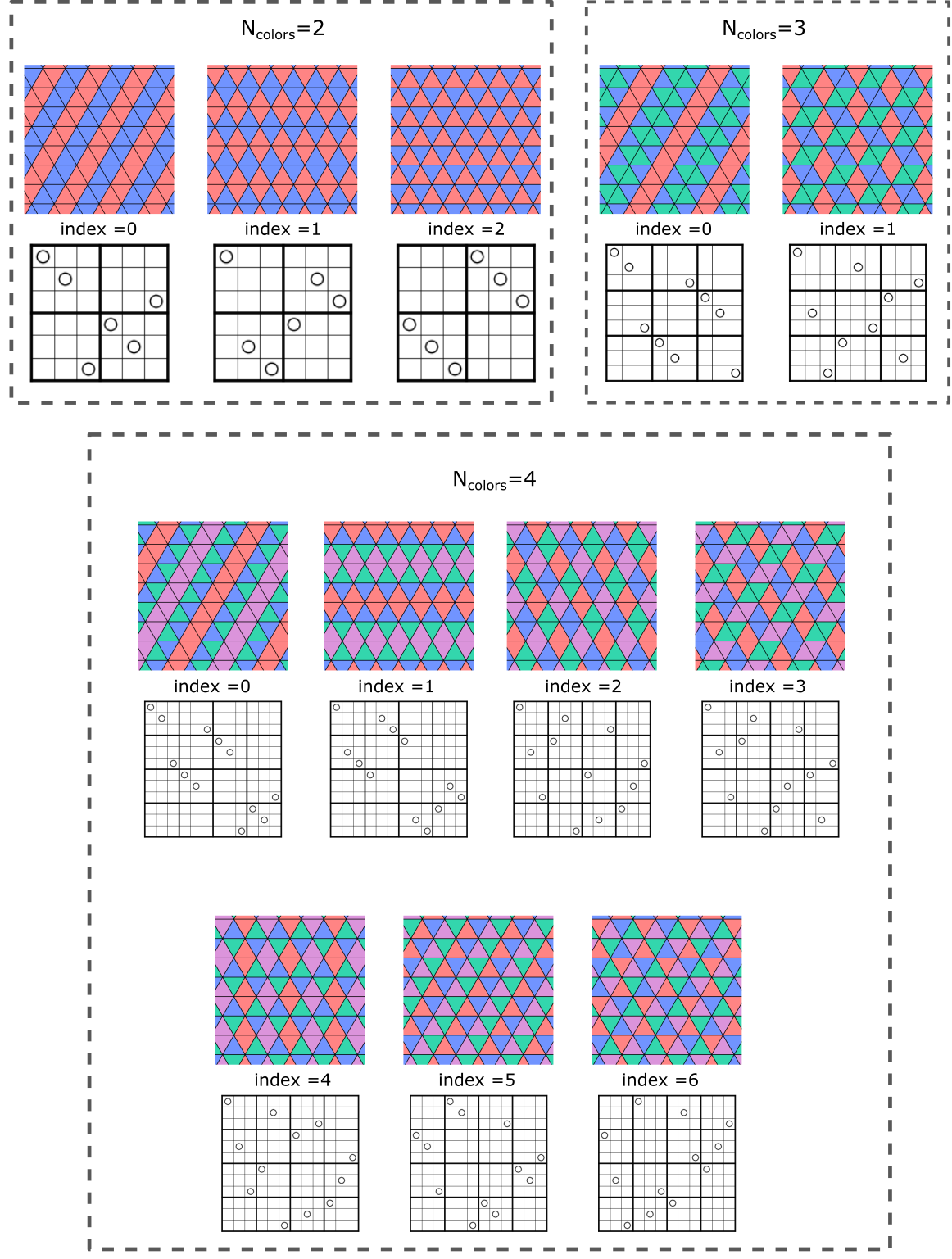}
    \caption{\textit{Color tilings}. Enumerated color tilings and their respective interaction matrices for $N_\mathrm{colors}=2-4$. The index numbers match those in Table~\ref{tab:my_label}}
    \label{fig:figS2}
\end{figure*}

\begin{figure*}[tbh]
    \centering
    \includegraphics[width=1\linewidth]{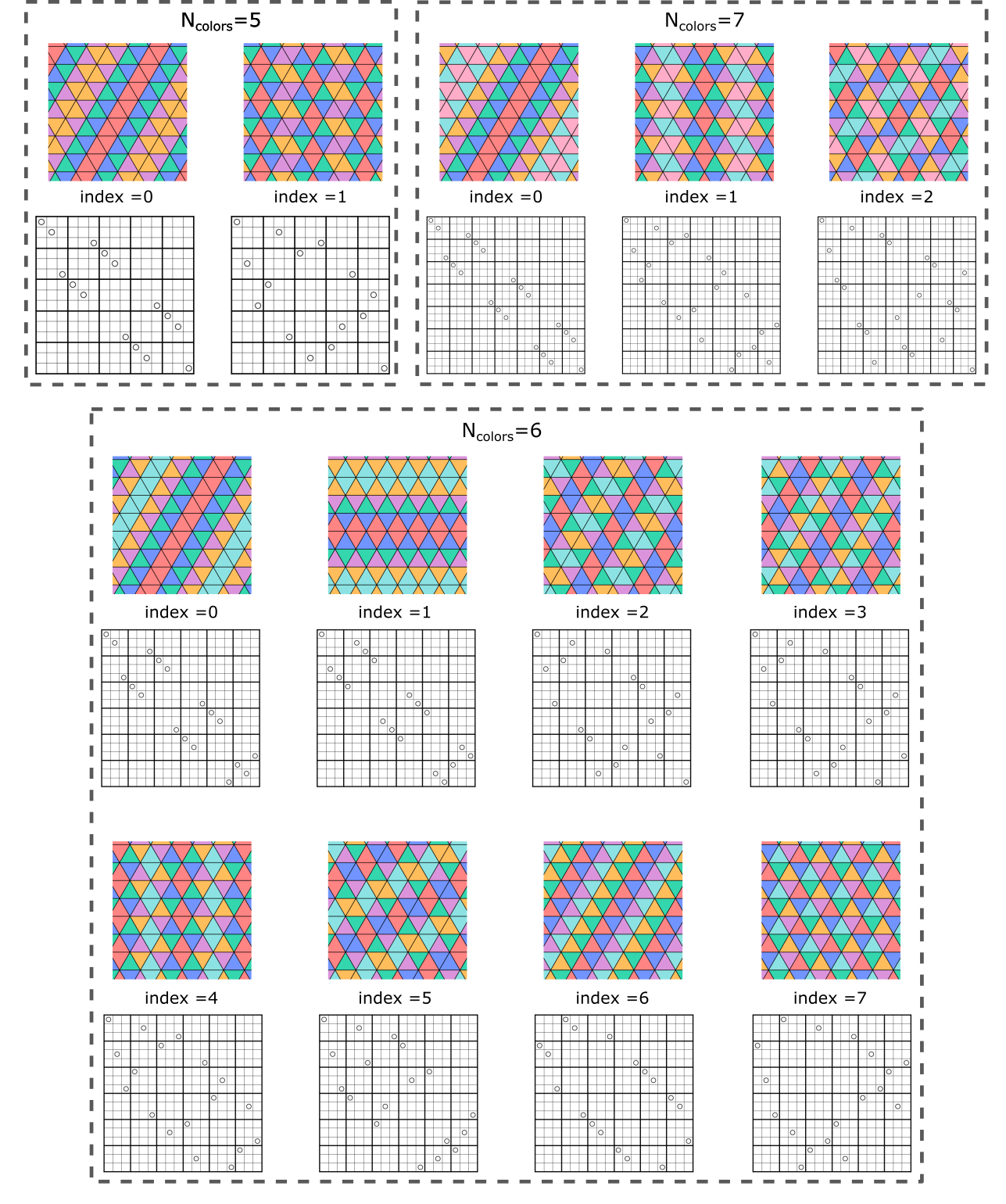}
    \caption{\textit{Color tilings}. Enumerated color tilings and their respective interaction matrices for $N_\mathrm{colors}=5-7$. The index numbers match those in Table~\ref{tab:my_label}}
    \label{fig:figS3}
\end{figure*}

\begin{figure*}[tbh]
    \centering
    \includegraphics[width=1\linewidth]{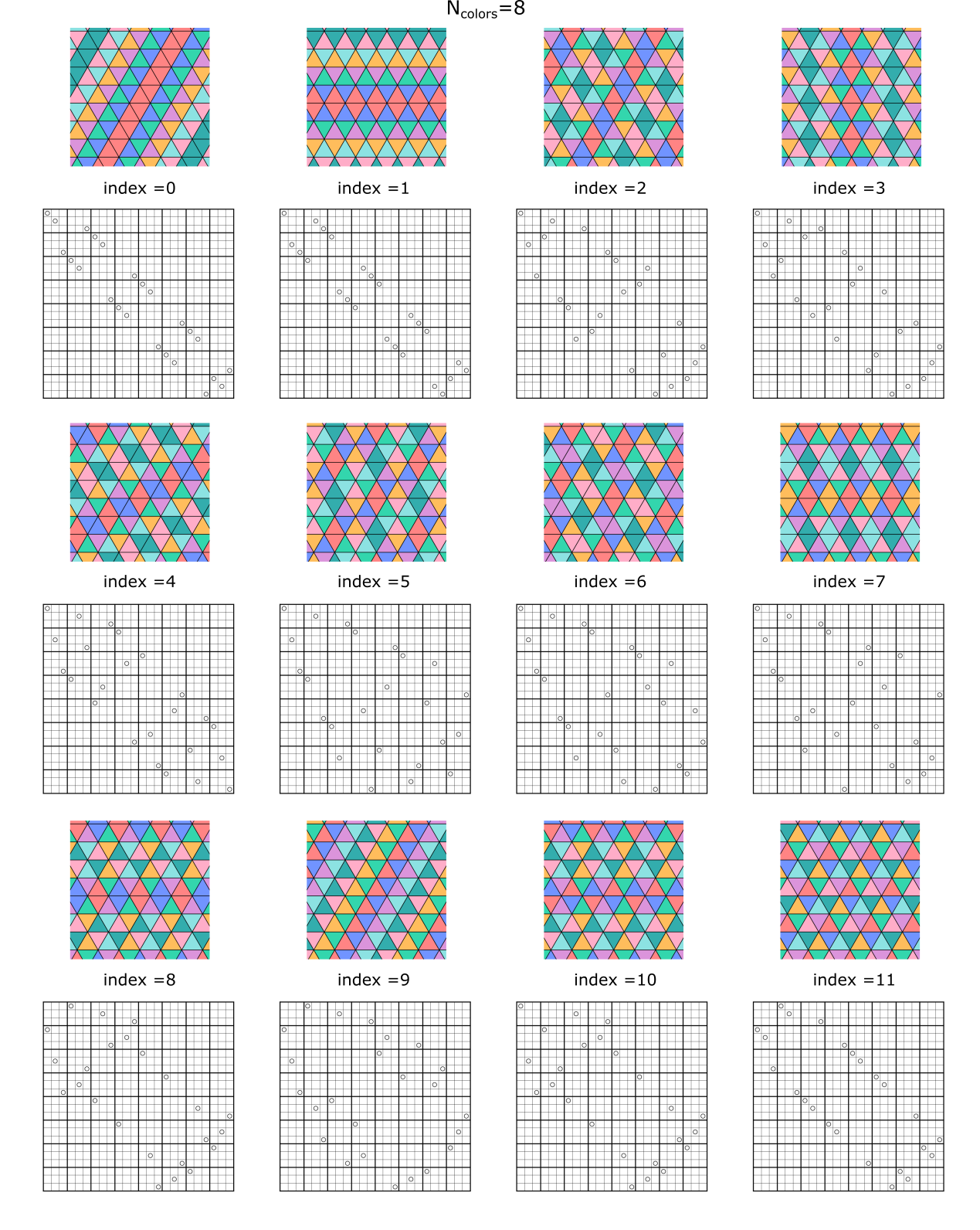}
    \caption{\textit{Color tilings}. Enumerated color tilings and their respective interaction matrices for $N_\mathrm{colors}=8$. The index numbers match those in Table~\ref{tab:my_label}}
    \label{fig:figS4}
\end{figure*}

\begin{figure*}[tbh]
    \centering
    \includegraphics[width=1\linewidth]{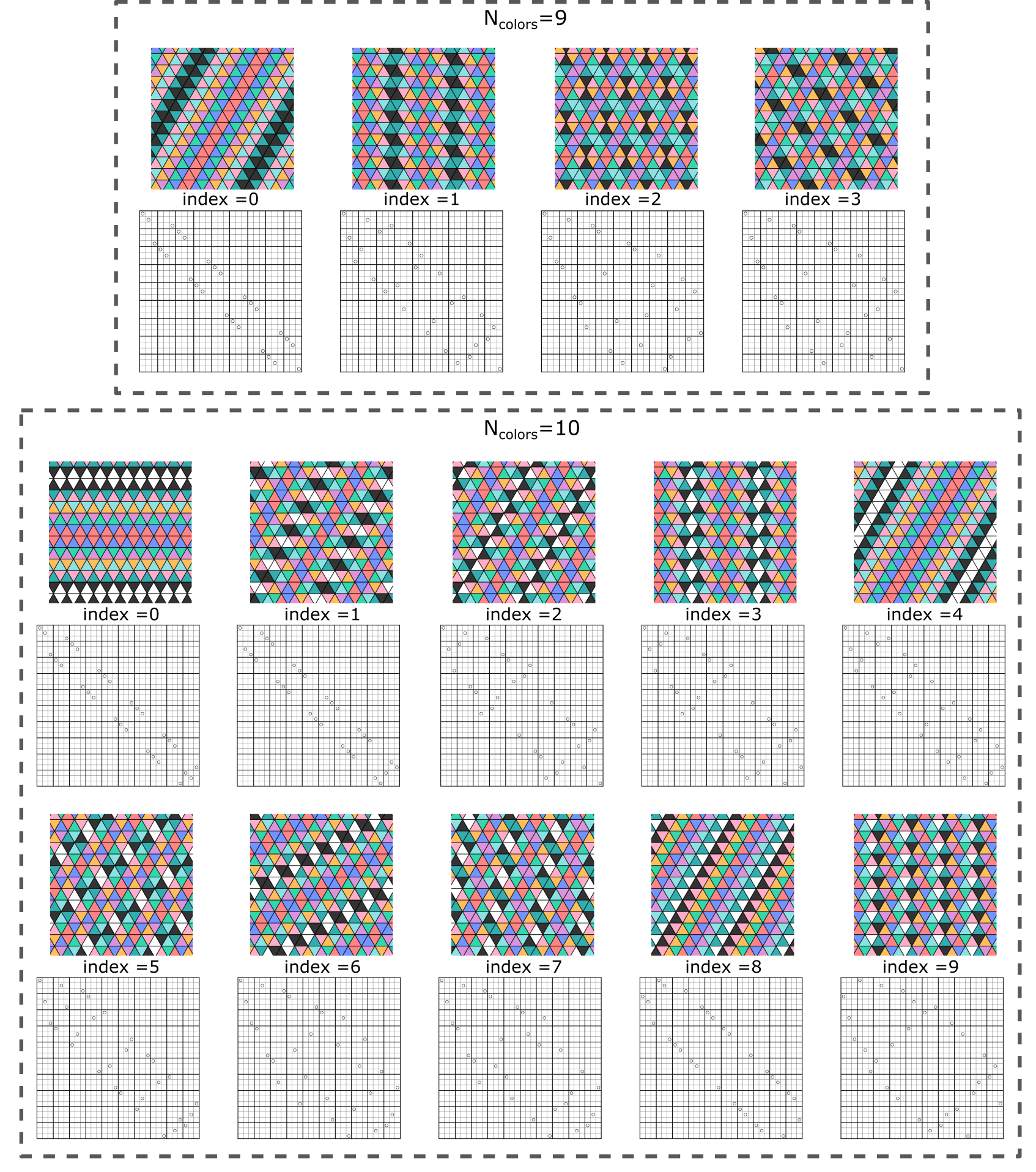}
    \caption{\textit{Color tilings}. Enumerated color tilings and their respective interaction matrices for $N_\mathrm{colors}=9-10$. The index numbers match those in Table~\ref{tab:my_label}}
    \label{fig:figS5}
\end{figure*}

Following these restrictions on the types of matrices and removing degenerate tilings, we have found all valid tilings that can be used to form tubules up to 10 colors. These tiling patterns and their interaction matrices are shown in figures~\ref{fig:figS2}-\ref{fig:figS5}. Another important property of these patterns is their primitive vectors, which are are listed in table~\ref{tab:my_label}.

\begin{table}[tbh]
\centering
    \begin{tabular}{c c c c}
         $N_\mathrm{colors}$ & index & $\textbf{a}_1$ &  $\textbf{a}_2$ \\ \hline \\
2 & 0 & $\textbf{m}$ & $-\textbf{m}+2\textbf{n}$ \\
2 & 1 & $\textbf{m}$ & $-\textbf{m}+2\textbf{n}$ \\
2 & 2 & $\textbf{m}$ & $\textbf{n}$ \\
3 & 0 & $\textbf{m}$ & $-\textbf{m}+3\textbf{n}$ \\
3 & 1 & $\textbf{m}+\textbf{n}$ & $-\textbf{m}+2\textbf{n}$ \\
4 & 0 & $\textbf{m}$ & $-2\textbf{m}+4\textbf{n}$ \\
4 & 1 & $\textbf{m}$ & $-2\textbf{m}+4\textbf{n}$ \\
4 & 2 & $2\textbf{m}$ & $2\textbf{n}$ \\
4 & 3 & $2\textbf{m}-\textbf{n}$ & $2\textbf{n}$ \\
4 & 4 & $2\textbf{m}$ & $-\textbf{m}+2\textbf{n}$ \\
4 & 5 & $2\textbf{m}$ & $\textbf{n}$ \\
4 & 6 & $2\textbf{m}$ & $2\textbf{n}$ \\
5 & 0 & $\textbf{m}$ & $-2\textbf{m}+5\textbf{n}$ \\
5 & 1 & $2\textbf{m}+\textbf{n}$ & $-\textbf{m}+2\textbf{n}$ \\
6 & 0 & $\textbf{m}$ & $-3\textbf{m}+6\textbf{n}$ \\
6 & 1 & $\textbf{m}$ & $-3\textbf{m}+6\textbf{n}$ \\
6 & 2 & $3\textbf{m}-\textbf{n}$ & $2\textbf{n}$ \\
6 & 3 & $3\textbf{m}-2\textbf{n}$ & $2\textbf{n}$ \\
6 & 4 & $3\textbf{m}$ & $-\textbf{m}+2\textbf{n}$ \\
6 & 5 & $\textbf{m}+\textbf{n}$ & $-3\textbf{m}+3\textbf{n}$ \\
6 & 6 & $\textbf{m}$ & $-\textbf{m}+3\textbf{n}$ \\
6 & 7 & $\textbf{m}+\textbf{n}$ & $-\textbf{m}+2\textbf{n}$ \\
7 & 0 & $\textbf{m}$ & $-3\textbf{m}+7\textbf{n}$ \\
7 & 1 & $3\textbf{m}+\textbf{n}$ & $-\textbf{m}+2\textbf{n}$ \\
7 & 2 & $3\textbf{m}-\textbf{n}$ & $-2\textbf{m}+3\textbf{n}$ \\
\\
\end{tabular}
\hspace{1em}
\begin{tabular}{c c c c}
$N_\mathrm{colors}$ & index & $\textbf{a}_1$ &  $\textbf{a}_2$ \\ \hline \\
8 & 0 & $\textbf{m}$ & $-4\textbf{m}+8\textbf{n}$ \\
8 & 1 & $\textbf{m}$ & $-4\textbf{m}+8\textbf{n}$ \\
8 & 2 & $2\textbf{m}$ & $-2\textbf{m}+4\textbf{n}$ \\
8 & 3 & $4\textbf{m}$ & $-\textbf{m}+2\textbf{n}$ \\
8 & 4 & $4\textbf{m}$ & $-\textbf{m}+2\textbf{n}$ \\
8 & 5 & $\textbf{m}+2\textbf{n}$ & $-2\textbf{m}+4\textbf{n}$ \\
8 & 6 & $3\textbf{m}-\textbf{n}$ & $-\textbf{m}+3\textbf{n}$ \\
8 & 7 & $2\textbf{m}$ & $-2\textbf{m}+4\textbf{n}$ \\
8 & 8 & $2\textbf{m}$ & $-\textbf{m}+4\textbf{n}$ \\
8 & 9 & $\textbf{m}$ & $-2\textbf{m}+4\textbf{n}$ \\
8 & 10 & $2\textbf{m}$ & $-2\textbf{m}+4\textbf{n}$ \\
8 & 11 & $2\textbf{m}$ & $2\textbf{n}$ \\
9 & 0 & $\textbf{m}$ & $-4\textbf{m}+9\textbf{n}$ \\
9 & 1 & $4\textbf{m}+\textbf{n}$ & $-\textbf{m}+2\textbf{n}$ \\
9 & 2 & $3\textbf{m}$ & $3\textbf{n}$ \\
9 & 3 & $3\textbf{m}-2\textbf{n}$ & $3\textbf{n}$ \\
10 & 0 & $\textbf{m}$ & $-5\textbf{m}+10\textbf{n}$ \\
10 & 1 & $\textbf{m}$ & $-5\textbf{m}+10\textbf{n}$ \\
10 & 2 & $2\textbf{m}$ & $-2\textbf{m}+5\textbf{n}$ \\
10 & 3 & $2\textbf{m}$ & $-3\textbf{m}+5\textbf{n}$ \\
10 & 4 & $5\textbf{m}$ & $-\textbf{m}+2\textbf{n}$ \\
10 & 5 & $5\textbf{m}$ & $-\textbf{m}+2\textbf{n}$ \\
10 & 6 & $\textbf{m}+2\textbf{n}$ & $-2\textbf{m}+3\textbf{n}$ \\
10 & 7 & $3\textbf{m}+\textbf{n}$ & $-\textbf{m}+3\textbf{n}$ \\
10 & 8 & $\textbf{m}$ & $-2\textbf{m}+5\textbf{n}$ \\
10 & 9 & $2\textbf{m}+\textbf{n}$ & $-\textbf{m}+2\textbf{n}$ \\
    \end{tabular}
    \caption{\textit{List of pattern vectors}. The primitive vectors are listed for all tilings found up to ten colors. $N_\mathrm{colors}$ denotes the number of colors in the tiling. The indices of these patterns matches those shown in figure~\ref{fig:figS2}-\ref{fig:figS5}. $\textbf{a}_{1,2}$ are the two primitive vectors where $\textbf{m}$ and $\textbf{n}$ are the vectors for a step along directions 3 and 2 in the lattice, as shown in figure~1c of the main text.}
    \label{tab:my_label}
\end{table}

\clearpage

\bibliography{main.bib}